%% file: eco_main.tex
\documentclass{article}

\usepackage{amsmath}
\usepackage{amsfonts}
\usepackage{amssymb}
\usepackage{graphicx}
\usepackage{algorithm,algpseudocode,nicefrac,graphicx,url}
\usepackage{subfigure}
\usepackage{sidecap}

\usepackage[left=3cm,right=3cm,top=3cm,bottom=3cm]{geometry}
\usepackage{breakcites}
\usepackage{authblk}
\usepackage{hyperref}

\newcommand{\Exp}{\mathbb{E}}
\newcommand{\vb}{\mathbf{b}}
\newcommand{\ve}{\mathbf{e}}
\newcommand{\vR}{\mathbf{R}}
\newcommand{\vA}{\mathbf{A}}
\newcommand{\vs}{\mathbf{s}}
\newcommand{\vx}{\mathbf{x}}
\newcommand{\vm}{\mathbf{m}}

\newcommand{\vH}{\mathbf{H}}
\newcommand{\vtheta}{\boldsymbol{\theta}}

\newtheorem{example}{Example}

\newtheorem{proposition}{Proposition}

\newcommand{\qed}{\hfill$\rule{2mm}{3mm}$}

\begin{document} 

\title{A General Framework for Evaluating \\Callout Mechanisms in Repeated Auctions}
\author[1]{Hossein Azari \thanks{\url{azari@claritymoney.com}}}
\author[2]{William D. Heavlin \thanks{\url{bheavlin@google.com}. This work was done while Hossein Azari was with Google research.}}
\author[3]{Hoda Heidari \thanks{\url{hoda@seas.upenn.edu}}}
\author[2]{Max Lin \thanks{\url{whlin@google.com}}}
\author[2]{Sonia Todorova \thanks{\url{todorova@google.com}}}
\affil[1]{Clarity Money Inc.}
\affil[2]{Google Inc.}
\affil[3]{University of Pennsylvania}

 \maketitle

\begin{abstract}
Motivated by online display ad exchanges, we study a setting in which an exchange repeatedly interacts with bidders who have \emph{quota}, making decisions about which subsets of bidders are called to participate in ad-slot-specific auctions. A bidder with quota cannot respond to more than a certain number of calls per second. In practice, random throttling is the principal solution by which these constraints are enforced. Given the repeated nature of the interaction with its bidders, the exchange has access to data containing information about each bidder's segments of interest. This information can be utilized to design smarter callout mechanisms --- with the potential of improving the exchange's long-term revenue. 
In this work, we present a general framework for evaluating and comparing the performance of various callout mechanisms using \emph{historical auction data only}. To measure the impact of a callout mechanism on long-term revenue, we propose a strategic model that captures the repeated interaction between the exchange and bidders. Our model leads us to two metrics for performance: immediate revenue impact and social welfare. Next we present an empirical framework for estimating these two metrics from historical data. For the baseline to compare against, we consider random throttling, as well as a greedy algorithm with certain theoretical guarantees. We propose several natural callout mechanisms and investigate them through our framework on both synthetic and real auction data. We characterize the conditions under which each heuristic performs well and show that, in addition to being computationally faster, in practice our heuristics consistently and significantly outperform the baselines.
\end{abstract}

\input{intro}

\input{related}

\input{setting}
\input{performance}
\input{heuristics}

\input{experiments}

\input{conclusion}

\bibliographystyle{apalike}
\bibliography{eco_bib}

\appendix
\input{technical}

\end{document}

%% file: intro.tex
\section{Introduction}
For online businesses advertising is a major source of monetization. Every day companies like Bing, Facebook, Google, and Yahoo run auctions --- billions of auctions --- to determine which advertising impressions to show. In particular, online display ad space is usually bought and sold through high-volume auction-based exchanges, of which AppNexus, Google's DoubleClick, and Microsoft Ad Exchange are examples.
In these online display ad exchanges, impressions are continuously received from publishers and auctioned off among real-time bidders. Economic theory holds that such auctions allocate resources efficiently. The auction also determines the price paid by the winner. Of this payment, a fixed percentage goes to the exchange and the remainder is paid to the publisher. These transactions constitute the two revenue streams of online advertising, those of ad exchanges and of publishers.

On the exchange side the process of running an auction usually consists of two steps: (1) Once a query arrives from the publisher side, the exchange calls a subset of buyers\footnote{While technically not the same, for simplicity in this work we use the terms ``buyer" and ``bidder" interchangeably.} to participate in the auction (the callout step). (2) Then, among the responding buyers the exchange runs an auction to determine the winner and price (the auction step). There are multiple reasons for the existence of the first step --- the focus of this work. First, a significant percentage 
 of bidders are limited by the number of calls per second they can respond to, their \emph{quota}~\cite{SelectiveCallouts,CEG,DJBW}. The exchange must protect these bidders from receiving more calls than their servers can handle\footnote{Beside such technological limitations, bidders may have financial constraints (see for example~\cite{Borgs,Dobzinski}) and/or specify volume limits to control exposure (see~\cite{Lahaie}).}. Furthermore, the exchange itself may need to limit the number of callouts sent to bidders to conserve its resources.
 
In practice, \emph{random quota throttling (RQT)} is the principal solution by which these constraints are enforced. At a high level RQT decides which buyers to call randomly and with probabilities proportional to the quota-per-seconds (qps). Given that the exchange interacts with its bidders repeatedly and over time, it has access to \emph{historical auction data} containing information about each bidder's segments of interest. By \emph{learning} this information the exchange can \emph{target} bidders more effectively. The combination of the learning and targeting is what we call a \emph{``callout mechanism"}. An ideal callout mechanism reduces resources when bidders are unlikely to be interested, and increases calls when bidders are more likely to perceive value.

Finding the optimal callout set is equivalent to solving the following optimization problem:  Subject to bidders' quota, which callouts should be sent to them for valuation and bidding so that the exchange's \emph{long-term revenue} is maximized? 
Finding the revenue-maximizing callout is computationally hard (Section~\ref{sec:setting}).
As a consequence, the exchange has to rely on heuristics and approximation callout mechanisms.
Different callout mechanisms can impact the long-term revenue differently, and the space of all possible callout mechanisms is extremely large.
It is therefore not feasible for the exchange to evaluate every mechanism by running an actual experiment\footnote{Experiments can be unpredictably costly. In addition, there are usually restrictions in place on buyers the exchange can run experiments on.}. 
This necessitates the design of a framework that utilizes \emph{historical data only} to compare the performance of various callout mechanisms.
This is the main contribution of the present work.
Rather than focusing on any particular mechanism, here we lay out a framework for evaluating any given callout mechanism in terms of its impact on long-term revenue.

The paper is organized as follows: 
We formalize the setting in Section \ref{sec:setting}.
In Section~\ref{sec:performance} we start by observing that different callout mechanisms can impact the long-term revenue differently, mainly for the following two reasons:
(1) Different mechanisms satisfy the quota differently.
(2) Bidders are strategic and adapt their response to the choice of the callout mechanism. 
Measuring the former can be readily done using historical data, however, to measure the latter we need to have a model for the way bidders adapt their behavior to a new callout mechanism. We propose in Section~\ref{subsec:game} a game-theoretic model that captures the repeated interaction between the exchange and its bidders.
This model motivates two performance metrics: 
\emph{immediate revenue impact} and \emph{social welfare}, both of which can be estimated from historical data (Section~\ref{subsec:estimators}).
To establish baselines for comparison, in Section \ref{sec:baselines} we consider two mechanisms:  RQT, as well as a greedy algorithm (GRA) for which theoretical guarantees have been established, albeit under certain restrictive assumptions. 
In Section \ref{sec:heuristics} we propose several natural callout heuristics. Finally in Section \ref{sec:experiments}, we demonstrate our empirical framework, measuring the performance of these callout mechanisms on both real-world and synthetic auction data.
We characterize the conditions under which each heuristic performs well and show that, in addition to being computationally faster, in practice our heuristics consistently and significantly outperform the baselines.

%% file: related.tex
\subsection{Related Work}\label{app:rel}


Microeconomics has a long line of research literature on auctions; see~\cite{Klem} for a survey. Repeated auctions, in which the auctioneer interacts with bidders multiple times, have received considerable attention~\cite{Bikhchandan,OW,Thomas}. In particular, the problem of \emph{pricing} in repeated auctions has been studied extensively~\cite{ARS'13,KL,BKRW,BHW,CGM,MM}.  That said, most of the previous work has not considered strategic buyers. Some consider random bidders~\cite{MM}, some study bidders who participate only in a single round~\cite{KL,BKRW,BHW,CGM}, and some focus on bidders who participate in multiple rounds of the auction, but are interested in acquiring exactly one copy of a single good~\cite{HKP}. In none of these cases do bidders react to the seller's past behavior to gain higher payoffs in future rounds. However, in many real-world applications, the same set of buyers interacts repeatedly with the same seller. There is empirical evidence suggesting that these buyers behave strategically, adapting to their history to induce better payoffs \cite{EO}. Indeed, a growing literature enumerates the various strategies buyers can follow in order to improve their payoff \cite{CDE,KL'4,JP,Lucier,GKP}. 

More recently, several studies have focused on the strategic aspect of bidders' behavior in repeated auctions and, in particular, on the problem of setting near-optimal prices. \cite{AV,KN} study the impact of Intertemporal Price Discrimination (IPD), i.e. conditioning the price on bidder's past behavior, and examine the conditions under which IPD becomes profitable. \cite{ARS'13,ARS'14} investigate repeated posted-price auctions with strategic buyers, and present adaptive pricing algorithms for auctioneers. With these algorithms the auctioneer's regret increases as a function of bidders' discount factors.

In this work, rather than optimizing reserve prices, we focus on improving the callout routine. Of course, targeting bidders to call to an auction has already been reduced to practice. Consider \emph{Selective callouts}, as implemented by Google's display ad exchange~\cite{SelectiveCallouts}: Unlike RQT, Google's selective callout algorithm (SCA) identifies the types of impressions a bidder values, thereby increasing the number of impressions the bidder bids on and reducing the number of callouts the bidder ignores. 

To our knowledge, \cite{CEG} is the first work to study the callout optimization problem from a purely theoretical perspective. The authors model callouts by an online recurrent Bayesian decision-making algorithm, one with bandwidth constraints and multiple performance guarantees. The optimization criteria considered in \cite{CEG} are different from ours --- here we consider long-term revenue as the principal criterion and assert that it is the most natural choice. Also, unlike \cite{CEG}, we study the effect of strategic buyer behavior on system-wide, long-term outcomes. Relatedly, \cite{Dughmi} investigates the conditions under which the expected revenue of a one-shot auction is a submodular function of the set of participating bidders. While these conditions are restrictive and do not often hold in practice, we adopt Dughmi's greedy algorithm as a baseline and compare other algorithms to it.

Our work is indirectly related to the large and growing body of research on budget constraints in online auctions. Papers in this literature can be divided into two main categories: (a) those concerned with the design and analysis of auction mechanisms with desirable properties --- truthfulness, optimal revenue, and so on --- for budget constrained bidders (see for example~\cite{Borgs,Hafalir,Dobzinski,BBGW}); and (b) those that present optimal or near-optimal bidding algorithms for such bidders (see for example~\cite{Chakrabarty,Archak}).

\textbf{Incentive Issues:} We close this section with a remark for readers familiar with the literature on incentive issues and truthfulness in mechanism design (see Chapter 23 in \cite{Mas-Colell} for an overview of the main results of this topic.)  While we are concerned with, and present shortly a model for, the bidders' strategic reactions to the choice of the callout mechanism and its impact on long-term revenue, we do not claim that a bidder is better off by \emph{bidding truthfully} when the system is in equilibrium. Indeed, in a setting as complicated as that with which we are dealing here --- in which bidders have complex strategy spaces and information structures --- the auction mechanism itself already fails to maintain incentive compatibility; see \cite{Borgs,Gonen} for related hardness and impossibility results. Rather than setting the ambitious goal of restoring bidding truthfulness, here we consider a model in which both the action space and the information structure are simplified (see Section \ref{subsec:game} for further details). In spite of the simplicity, the analysis of our model provides us with important insights about the performance of callout mechanisms.

%

%% file: setting.tex
\section{Setting and Preliminaries}\label{sec:setting}
Let $B = \{1,2,\cdots,n\}$ denote the set of bidders active in the exchange. Each bidder $i \in B$ has a quota-per-second constraint denoted by $q_i>0$. This quantity is known to the exchange and specifies the number of auctions bidder $i$ can be called to per second.
Consider a particular time interval of length one second, and assume that during this period the exchange receives a sequence of ad slots $A = \{a_t\}_{t=1}^T$ in an online fashion. 
Let $v^t_i$ denote the value (or ROI) of ad slot $a_t$ to bidder $i$. The exchange does not know the valuations in advance, but can learn about them through the bids. Let $b^t_{i}$ specify bidder $i$'s bid for the ad slot $a_t$. 

\begin{figure}[h!]
\centering 
\includegraphics[width=0.6\textwidth]{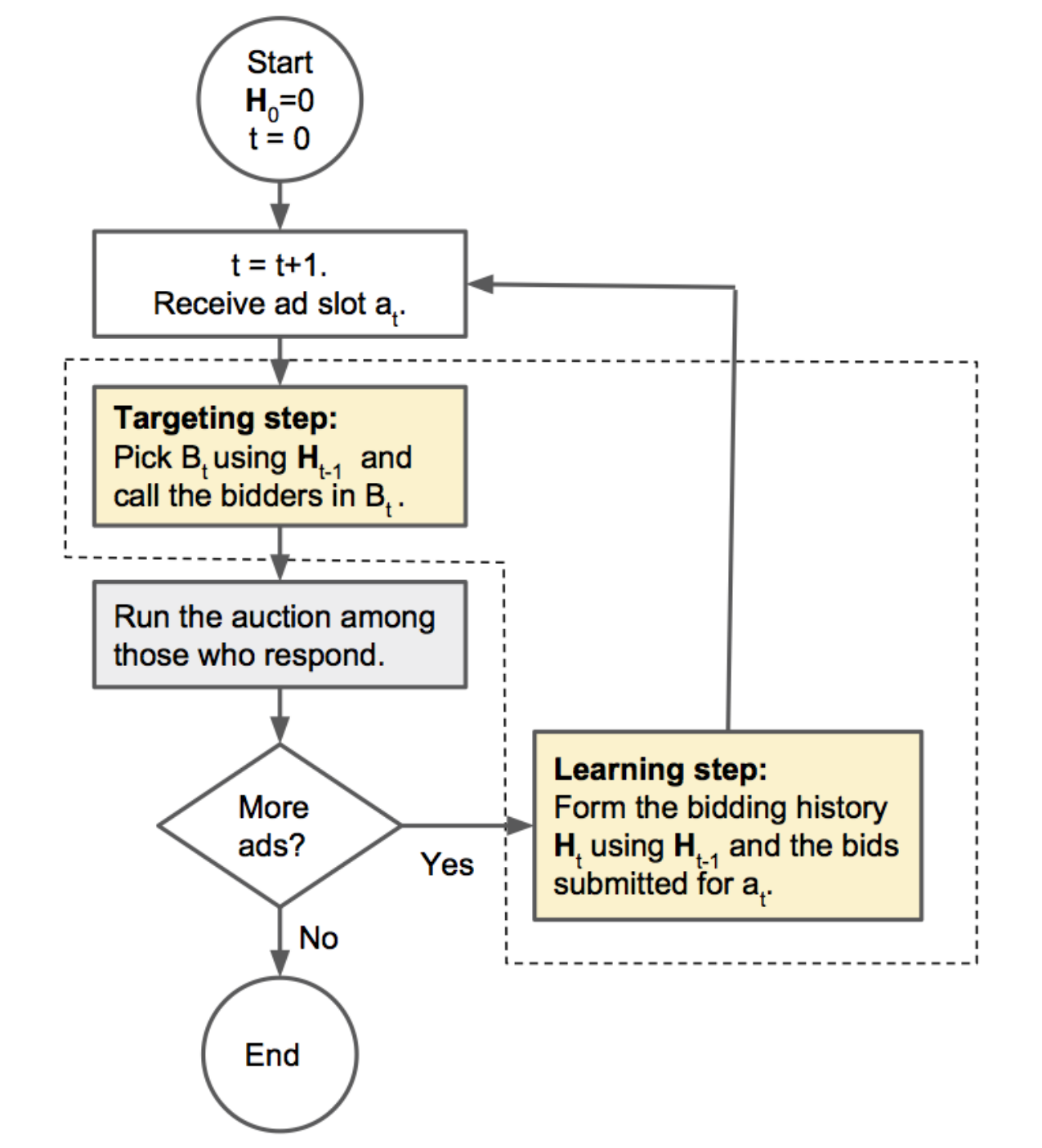}
\caption{The high-level flowchart of the system. The two main building blocks of a callout mechanism are the components for \textit{learning} and \textit{targeting}. We treat the auction mechanism as fixed and given throughout.} \label{fig:flowchart}
\end{figure}

At time $t=1,2,\cdots$ when ad slot $a_t$ arrives, the exchange must choose a subset $B_t \subseteq B$ to call for participation in the auction for $a_t$ while respecting all the quota-induced constraints.  A \textit{callout mechanism/algorithm} is any logic used to decide which subset of bidders to call at each time step $t$ using the history of bidding behavior observed up to time $t$. More precisely, let the matrix $\vH_t$ denote the bidding history observed by the exchange up to time $t$. Given $\vH_{t-1}$ as the input a callout mechanism selects $B_t$ for every $t=1,2,\cdots$. 
Once called, each bidder $i \in B_t$ decides whether to participate in the auction. The auction mechanism then specifies the winner (if any) and price. The recent bids along with $\vH_{t-1}$ are then used to set $\vH_{t}$. Figure \ref{fig:flowchart} 
illustrates the flow of a typical callout mechanism.

Throughout, we treat the auction step as given and fixed; we assert no control over how the auction is run or how its parameters, e.g. the reserve price, are tuned. While much of our work extends readily to other auctions, unless otherwise specified we assume the auction mechanism is second price with reserve. Let $r_t$ be the (given) reserve price for auction $t$. Bidders do not observe $r_t$ before submitting their bids. 


For the exchange, the ultimate goal is to maximize the \emph{long-term revenue}. Therefore we define
the performance of a callout mechanism by its impact on long-term revenue. Long-term revenue is simply the (possibly discounted) sum of the revenue earned at each time step $t=1,2,\cdots$. We denote the discount factor by $\delta$ ($0 \leq \delta \leq 1$). 
The optimal callout mechanism is the one that maximizes the long-term revenue.
We conclude this section by demonstrating that identifying the optimal callout is computationally hard. And this remains true even if we assume bids are sampled from fixed distributions, the exchange knows these distributions ahead of time, and can accurately predict the sequence of ad slots it receives.
\begin{proposition}\label{prop:hardness}
Suppose when called to an auction for item $t$, bidder $i$ samples their bid, $b^t_{i}$, from a distribution $D^t_i$ ($i=1,\cdots,n$ and $t=1,\cdots,T$).  Let variable $x_{i,t}$ be the indicator of whether bidder $i$ is called to the auction for $a_t$. Solving the following optimization problem is NP-hard:
\begin{eqnarray*}
\max && \sum_{t=1}^T \delta^t \Exp R(x_{1,t},\cdots,x_{n,t})\\
\text{s.t. } &&\forall i: \sum_{t=1}^T x_{i,t} \leq q_i \\
&& \forall i,t: x_{i,t} \in \{0,1\}
\end{eqnarray*} 
where the random variable $R(x_{1,t},\cdots,x_{n,t})$ denotes the revenue of auction $t$ with participants $\{i \vert x_{i,t}=1\}$, and the expectation is taken with respect to $D^t_i$'s.
\end{proposition}
We establish this by showing that the above class of problems includes a known NP-hard problem---maximum 3-dimensional matching---as a special case. Omitted proofs and other technical material can be found in Appendix~\ref{app:tech}.

Proposition~\ref{prop:hardness} motivates our approach in this work. Consider the hypothetical world in which we have access to a perfectly accurate \emph{bid forecasting} model. In this setting one natural proposal is to use this model to forecast the bids, then call the two bidders with the highest bids in every auction. We emphasize that in practice this is simply not possible: Even the best forecasting model can only forecast a \textit{bidding distribution}, not a particular number. Furthermore, even if we assume this distribution is fully compliant with the true bidding behavior, according to Propostion~\ref{prop:hardness} finding the optimal callout set is computationally hard. This implies that the exchange has to rely on heuristics and approximation algorithms to construct the callout set. In this work we present a data-driven approach for evaluating and comparing various callout algorithms in terms of their average impact on long-term revenue.

%% file: performance.tex
\section{Performance}\label{sec:performance}
We start by observing---via a series of examples in Appendix~\ref{app:ex}---that different callout mechanisms result in different levels of long-term revenue for the following two reasons:
(1) Different mechanisms satisfy quota differently.
(2) Bidders are strategic and adapt their response to the choice of the callout mechanism. 
Measuring the first type of impact (illustrated in Example~\ref{ex:1}) is readily possible using historical data --- one simply needs to run each callout mechanism on the data and observe the change in revenue. As illustrated by Example~\ref{ex:1} by selling previously unfilled impressions, a smart callout mechanism can increase the exchange's revenue, and at the same time maintain, and perhaps improve, bidders' utilities. 

In order to improve long-term revenue, however, it does not suffice for the exchange to find a callout mechanism with high revenue performance on historical data. In real world bidders are strategic and adapt their response to the choice of the callout mechanism. 
In Example \ref{ex:2} for instance, callout mechanism (2) does not result in selling previously unfilled part of the inventory. Rather, it merely increases the competition and as a result, the price for item 2. While this does not hurt the revenue immediately, it reduces the utility that the bidders (in particular, bidder 3) earn. Bidders potentially react to this change in their payoffs in future rounds. To quantify this type of impact on long-term revenue we need to have a \emph{model} for bidders' reactions to the choice of the callout mechanism.

\subsection{A Two-Stage Game Formulation}\label{subsec:game}
There are usually many options (i.e exchanges) available to bidders to choose from (i.e. participate in).
We make the following simple, natural assumption about the bidder's reaction to the choice of the callout mechanism in one particular exchange: Bidder seek to maximize their \emph{utility} (ROI), that is, they always choose to participate in the exchange that provides them with the highest utility (value minus cost). In what follows, we make two simplifying assumptions: First we limit the bidder's action space to the choice of which exchange to \emph{participate} in. Second, we assume by participating in an exchange, all that the bidder can observe is their utility from that exchange. We acknowledge that compared to real-world, these assumptions drastically simplify the action space and the information structure for both the bidders and the exchange. Nonetheless as the following analysis shows, this simplified model suffices to capture some of the most important aspects of the interaction between the exchange and bidders. 

We follow the standard modeling approach in microeconomic theory for repeated interactions among strategic agents (see~\cite{MS}).
Consider the following two-stage game between two players, the exchange and one bidder. The action space for the exchange consists of all possible callout mechanisms, denoted by the set $E$. The action space for the bidder consists of two actions: participating or taking the outside option.
At the beginning of the first stage the exchange commits to a callout mechanism $e \in E$. (Note that our analysis does not rely on a particular choice of $e$.) Next, the bidder, without the knowledge of $e$, decides whether to participate in the exchange in the first round. If they do, then their expected utility is equal to
$$u^e = v^e - c^e $$
where $v^e$ is the bidder's average valuation for the items they win and $c^e$ is the average cost they pay for those items under callout mechanism $e$. If the bidder chooses to participate, the expected revenue that the exchange receives equals $c^e$. If they do not participate in the auction, they take their outside option, which provides them with utility $u$ and the exchange with an incremental revenue of $0$.

In the second stage of the game, which captures the future interaction of the bidder with the exchange, the bidder again chooses whether to participate or not. This time, however, the bidder is aware of the utility they earns from participation --- if they chose to participate in the first stage. Note that this is in line with the anecdotal evidence suggesting that bidders do in practice run experiments to estimate the utility they can earn from different exchanges and adjust their rates of participation accordingly. Utilities for this stage are defined similarly to the first stage. Denote by $\delta$ the players' discount factor for the future (i.e. the second stage) payoffs.


\begin{proposition}\label{prop:eq}
Among all callout mechanisms $e \in E$ for which $u^e \geq u$, let $e^*$ be the one for which $c^e$ is maximized.  
If $\frac{\max_e c^e}{c^{e^*}} -1 \leq \delta$, the following strategy profile is the unique sub-game perfect equilibrium of the above game: The exchange chooses $e^*$, and the bidder chooses to participate in each stage if and only if according to their beliefs at that stage $\Exp u^e \geq u$.
\end{proposition}
The above proposition suggests that when players value their future utilities --- or more precisely, when the discount factor $\delta$ is large enough --- the ideal callout mechanism increases the immediate revenue as much as possible (i.e. maximizes $c^e$)  while providing the bidder with a utility level at least as large as what they can earn from their outside option (i.e. maintaining the constraint $u^e \geq u$).
In other words, when choosing a callout mechanism, the exchange faces a trade-off between immediate- and long-term revenue: Of course, for a given callout mechanism, increasing its callouts can boost the revenue. However, unless the exchange induces sufficiently high value for bidders, such increases in callouts ultimately discourage their participation in future rounds --- they find their outside option more profitable. This in turn translates into less revenue for our exchange in the long run. 

\subsection{Performance Metrics and Estimators}\label{subsec:estimators}
The argument above leads us to two metrics for evaluating the performance of a callout mechanism: immediate revenue impact and social welfare. Next we propose ways for estimating these metrics, $c^e$, $u^e$, as well as the outside option utility $u$, from historical bidding data. Throughout we assume we know and can therefore simulate the auction mechanism on any given set of bids.
 
More precisely, suppose we have access to the bidding data over a period of length $S$ for $n$ bidders (i.e. $b^t_i$ for all $t=1,\cdots,S$ and $i=1,\cdots,n$). 
We will need the following notation: 
Let $\tilde{c}_{i}$ be the average cost bidder $i$ pays for the items they win over time period $t=1,\cdots,S$---before a new callout mechanism is implemented. Similarly, let $\tilde{b}_{i}$ be the average bid submitted by bidder $i$ for the items they win over this time period.
Let $\tilde{c}_{i}$ be the average cost bidder $i$ pays for the items they win over time period $t=1,\cdots,S$---under callout mechanism $e$. Note that this can be easily computed by simulating the auction mechanism on the historical bidding data and the new callout set. Similarly, let $\tilde{b}_{i}$ be the average bid submitted by bidder $i$ for the item they wins over this time period---under callout mechanism $e$.
\paragraph{Immediate revenue impact} The following is an unbiased estimator of $c^e$:
$\bar{c}^e = \frac{1}{n} \sum_{i=1}^n  \tilde{c}^e_{i}.$
\paragraph{Social welfare} To estimate the social welfare, we need some proxy of the bidders' valuations. Since we do not have access to actual valuations, for practical reasons we are constrained to rely on bids as a proxy for value. In our setting, the assumption of bid as a proxy for valuation is relatively benign: any bias in measuring the utility of winning auctions in one exchange is likely the same bias for winning auctions in any other exchange. Further, the choice of bid-as-value enables bid-minus-cost as the residual value for the buyer, one that is visible both to each buyer and to the exchange. In that sense, bid-minus-cost represents the good-faith estimate of the residual value, one that the exchange can actively work to preserve over the set of buyers.
Assuming that a bidder's average bid reflects their true valuation, the following is an unbiased estimator for $u^e$:
$\bar{u}^e = \frac{1}{n} \sum_{i=1}^n  (\tilde{b}^e_{i} - \tilde{c}^e_{i}).$
\paragraph{Outside option utility}  The following is an estimator for $u$:
$\bar{u}^e = \frac{1}{n} \sum_{i=1}^n (\tilde{b}_i - \tilde{c}_i).$
We argue that the above is an unbiased estimator assuming that bidders are rational and participate at positive rates in both the exchange and their outside option.
Here is why: 
Consider a strategic bidder who participate at positive rates in both our exchange and the outside option. This means that both of these options provide them with equal utilities on average, otherwise, they would have been better of by only participating in the higher paying exchange. 

%% file: heuristics.tex
\section{Our Callout Algorithms}\label{sec:heuristics}

\subsection{Baselines}\label{sec:baselines}
We propose two baselines, RQT and GRA, against which we compare other callout mechanisms\footnote{Note that SCA~\cite{SelectiveCallouts} is not among our baselines, because the details of the algorithm have not been published.}.

\paragraph{Random Quota Throttling (RQT)}
As a naive comparison baseline, we consider random quota throttling, RQT: we drop each bidder with a fixed probability $p$. 

\paragraph{The Greedy Algorithm (GRA)}
In settings where the auction revenue is monotone and submodular\footnote{This is for example the case for revenue maximizing auction mechanisms in  matroid markets with bidders whose bids are drawn independently (see \cite{Dughmi} for the details).} 
a simple greedy algorithm, which greedily adds bidders by their marginal impact on revenue, is guaranteed to obtain revenue at least as large as $(1-1/e)$ of the (one-shot) optimal solution~\cite{Dughmi}. The details of the algorithm can be found in Appendix~\ref{app:other}.

\subsection{Our Heuristics}
Throughout this section we focus on callout mechanisms with two natural properties, \textit{symmetricity} and \textit{myopicity}. 
We call a callout heuristic \textit{symmetric} if two bidders with exactly the same bidding history up to time $t$ have the same chance of being called to the auction of $a_t$. This basic property must be satisfied to ensure that the mechanism is fair. 
A \textit{myopic} callout heuristic disregards the effect of today's targeting on the future bidding behavior. 
The targeting in all of our algorithm is done via \emph{thresholding}: the algorithm calculates a score vector $\vs_t$ from the (relevant part of the) history vector $\vH_t$ for all bidders; by designating a suitable thresholds $\vtheta$, it selects which bidders to call out. See Algorithm \ref{alg:sym} for details. Next we present several natural choices for the metric $m(.)$ and specify how to update the scores with it.


\begin{algorithm}
\caption{A Myopic Symmetric Thresholding Algorithm}
\label{alg:sym}
\begin{algorithmic}[1]
\State \textbf{Input:} metric $\vm(.)$ and threshold $\theta$.
\State Start with $t = 0$ and equal scores for all bidders ($\vs_0 = \mathbf{0}$).
\While{there exist more ad slots}
\State $t=t+1$.
\State Receive ad slot $a_t$.
\State Set $B_t = {i: s^i_{t-1} \geq \theta}$. 
\State Run the auction for $a_t$ among bidders in $B_t$.
\State Update $\vm_{t}$ by including the bids for $a_t$.
\State Update $\vs_t$ using $\vs_{t-1}$ and $\vm_{t}$.
\EndWhile
\end{algorithmic}
\end{algorithm}

We end this section with a remark: we restrict ourselves to thresholding mechanisms---as opposed to the broader class of \emph{randomized} bidder targeting---because the expected revenue earned from any randomized targeting can be written as the weighted sum of revenue earned by different $\vtheta$ vectors. This sum is maximized when all the weight is put on a single (best) threshold vector $\vtheta$.

\subsection{A Linear Heuristic}\label{sec:linear}
Consider an auction with $n$ bidders and let $b_i$ denote the bid of bidder $i$. Let $\vb = (b_{(1)},b_{(2)},\cdots ,b_{(n)})$ denote the ordered bid vector where the subscript $(j)$ denotes the $j$-th order statistic: $b_{(1)} \geq b_{(2)} \geq \cdots \geq b_{(n)}$.
For any $S \subseteq B$, let $R(S)$ denote the revenue that the exchange earns if it calls all the bidders in $S$ to the auction. We want to attribute to each bidder $i \in S$ part of the total revenue, which we denote by $R_i$, such that the attribution satisfies the following properties for any set $S$:\footnote{The axioms introduced here bear similarity to those leading to Shapley values in cooperative game theory~\cite{Shapley,Neyman}, hence the naming.}
\begin{enumerate}
\item \textit{Symmetry:} bidders with equal bids are attributed the same revenue:  $i,j \in S$ and $b_i=b_j \Rightarrow R_i = R_j$.
\item \textit{Linearity:} $\vR_S = \vA \vb_S $ for some fixed matrix $\vA$. Throughout, for any vector $\vx$ and any set $S$, $\vx_S$ is a vector that is equal to 0 on any component $i \not\in S$ and equal to $\vx$ everywhere else.
\item \textit{Conservation of Revenue:} the sum of attributions equals the total revenue:  $\sum_{i\in S} R_i = R(S)$.
\end{enumerate}


\begin{proposition}\label{prop:ShA}
For a second-price auction properties 1--3 uniquely identify $\vA$.
\end{proposition}

Shapley's Linear Heuristic (ShA) works by computing and thresholding on the average value of the above metric for each bidder.
We next argue that the above heuristic estimates the expected counterfactual revenue impact of adding a new bidder to an auction with respect to a particular distribution.
Let $S$ denote the subset of bidders called to the auction, but without bidder $i$. Consider two almost-duplicate worlds, one with bidders $S \cup {i}$ called to the auctions (the observable) and the other without the bidder in question, $S$ (not observable, i.e. counterfactual). If everybody participates, the impact on revenue of the intervention --- including $i$ --- is $\left(R(S \cup {i}) - R(S)\right)$. However, the set of bidders who actually end up participating in the auction is a random variable. Suppose the probability of the subset $T \subseteq S$ of bidders ending up participating is $\Pr(T)$. Then we can write the expected revenue impact of adding $i$ to $S$ as follows:
\begin{equation*}
R(S \cup {i})-R(S) = \sum_{T \subseteq S} \Pr(T) \left(R(T \cup {i}) - R(T)\right)
\end{equation*}
It is easy to see that the above is exactly equivalent to the calculation in Proposition \ref{prop:ShA} if
$\Pr(T) = \frac{|T|!(|S| - |T| -  1)!}{|S|!}$.
The above distribution is uniquely imposed due to the symmetry property\footnote{
As a future generalization, one can discard this property and consider a more general linear approach in which $ A^{(P)} \times \vec{e}_k = \vec{p}_k$, 
where $\vec{p}_k= (p_{k1},..,p_{kk},0,..,0),$ $\sum^k_{l=1} p_{kl} = 1$, and $p_{kl} \geq 0$.}.

\subsection{Non-linear Heuristics}\label{sec:non-linear}
We now turn to propose and discuss the following (nonlinear) heuristics. 
\paragraph{History of bidding above reserve (BAR)} We call bidder $i$ to an auction if the number of times she has in the past bid above the reserve price  for similar items exceeds some threshold $\theta_i$. Obviously, as we increase $\theta_i$, the number of bidders called to the auction decreases. As we will see in Section \ref{sec:experiments}, the performance of this heuristic depends heavily on the reserve price setting algorithm. The more accurate this algorithm is --- in predicting the auction winner, in predicting the winner's bid --- the better this heuristic performs. In the ideal case where the pricing algorithm can predict exactly the winner's bid, the BAR heuristic maximizes the revenue: we only need to call to the auction the person who is willing to pay the most and set the reserve price to a level just below her bid.
Conversely, consider the extreme case when the reserve price is 0 and therefore contains no information about bidder interest: the bid-above-reserve metric is equal for all bidders, and the heuristic BAR therefore performs poorly.

\paragraph{History of winning (WIN)}
We call bidder $i$ to an auction if the average number of times she has won in the past for similar items exceeds some threshold $\theta_i$. This algorithm performs well in the absence of an accurate reserve price setting algorithm for the following reason: The number of times a bidder wins in a segment indicates how interested she is in similar impressions. So by calling these interested bidders, competition (and, therefore, the second price) increases. The problem with this approach is that when multiple bidders have equal interest in an impression segment. Instead of splitting the impressions among them, this heuristic calls them simultaneously, driving competition up and dissipating bidders' resources. To the extent that the price setting algorithm is accurate, WIN is wasteful.
In addition to the above drawback, a bidder may not win often in a segment, but still succeed in setting a high second price for the winner. The WIN heuristic ignores this effect and does not call such price-pressuring bidders to the auction.

\paragraph{Total spend (SPD)}
We call bidder $i$ to an auction if her total spend for similar items so far exceeds some threshold $\theta_i$.
This heuristic can be thought of as an extension of WIN, one weighted not only by how many times a bidder wins in a segment, but also by how much she spends upon winning.

\paragraph{Average ranking (RNK)}
We call bidder $i$ to an auction if her average rank in past auctions for similar items lies below some threshold $\theta_i$.
This heuristic can be thought of as a generalized and smoothed version of WIN. With this heuristic the winner (i.e. the first ranked bidder) is not the only one who receives credit. Rather, every bidder increases her score proportional to the placement of where her bids stand relative to others.

\paragraph{Total bid (BID)}
We call bidder $i$ to an auction if her total past bids for similar items exceeds some threshold $\theta_i$.
The problem with this heuristic is the following: Consider a bidder who bids low most of the time, but every once in a while submits an unreasonably high bid to raise their average. This heuristic cannot distinguish this bidder from one that consistently submits reasonably high bids. 

\paragraph{Total attributed revenue (RVC)}
We call bidder $i$ to an auction if her total attributed revenue for similar items exceeds some threshold $\theta_i$. Note that a bidder's revenue impact manifests not only when she directly provides the winning bid, but also indirectly when she influences the price of any other winners.
The problem with this heuristic is that it completely disregards the role bidders other than the first and second-highest could have played in the auction. When the number of repetitions is not high, we expect ShA to outperform this heuristic. As the number of repetitions increase, this heuristic converges to ShA.





%
%
Table \ref{tab:rt} compares the running time of our algorithms and baselines with one another. Shortly in Section \ref{sec:experiments} we see not only that our heuristics are faster, but also that they outperform both baselines. One heuristic that deserves particular attention is ShA. ShA does not suffer from the problems pointed out for non-linear heuristics above; we therefore expect it to outperform them in practice. In Section \ref{sec:experiments} we see that this is indeed the case.


%% file: experiments.tex
\section{Experiments}\label{sec:experiments}
In this section we demonstrate our framework on both synthetic and real-world auction data. By simulating each mechanism on such data, we estimate its immediate revenue impact and social welfare impact using the estimators proposed in Section \ref{subsec:estimators}, and compare them with the baselines in Section \ref{sec:baselines}. As predicted earlier, we see that most of our heuristics consistently outperform the baselines.

At a high level, our simulator receives the auction data and processes it one auction at a time using the heuristic specified. For any given item, the simulator decides which subset of bidders to call to the auction for that item (by setting the threshold value), simulates the auction mechanism among those bidders, and finally calculates the revenue and social welfare of the auction. By changing the threshold value $\theta$, the percentage of called-out bidders varies. That allows us to obtain a range of values for the performance metrics for each heuristic as a function of the percentage of bidders called out.

\begin{figure}[h!] 
\centering 
\includegraphics[width=0.7\textwidth]{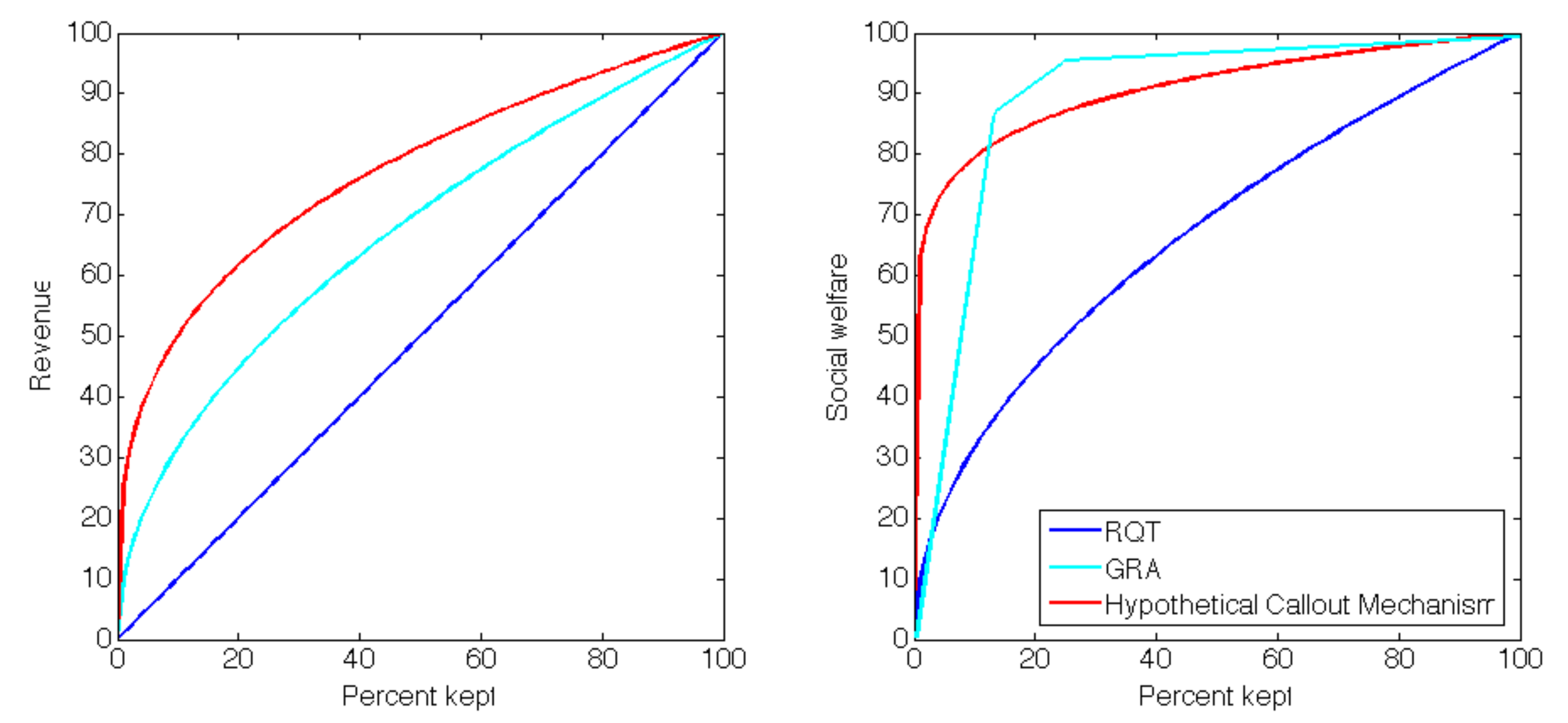}
\caption{A good callout mechanism (red) must exceed the baselines RQT (blue) and GRA (cyan) in terms of revenue (left panel) while maintaining social welfare at least as large as RQT (right panel).} 
\label{fig:baseline}
\end{figure}

In our simulations, we assume the qps rates are constant across bidders ($q_i =c$ for all $i$ and some constant $c$). This not only simplifies the simulation, but also, by enlarging the number of potential buyers in each auction, it implicitly increases the scale of each simulation. In practice, when different bidders have different qps's, one can designate different threshold values for each one of them; these thresholds can be set to guarantee no bidder is called out to more than their qps.
More importantly, the above choice allows us to see how each mechanism's performance evolves as the percent of bidders it can keep in the auction increases (i.e as we vary $p$).
Based on the argument in Section~\ref{sec:performance}, a callout mechanism outperforms the baselines if: 
(1) By calling the same percentage $p$ of bidders to auctions, it results in revenue higher than both RQT and GRA. 
(2) By calling the same percentage $p$ of bidders to auctions, it results in social welfare at least as large as RQT. 
For example, in Figure \ref{fig:baseline} the hypothetical callout mechanism outperforms both baselines.

\subsection{Datasets}
\paragraph{Synthetic auction data} 
In this dataset each bidder's bid is sampled from a fixed distribution specific to that bidder. 
We generate the data as follows: We assume there is a total of $T$ items that arrive at time steps $1,2,\ldots,T$. We have $n$ bidders and each one of them samples their bid from a \textit{log-normal} distribution with a fixed but bidder-specific median and variance. Note that the assumption that bids follow a log-normal distribution is standard in the literature and is backed by multiple empirical studies. See for instance~\cite{wilson1998sequential,xiao2009optimal,ostrovsky2011reserve}. For each bidder the mean bid and variance is sampled from the \textit{uniform} distribution with support on $[0,\mu]$ and $[0,\sigma]$, respectively. For simplicity, we assume the reserve price is \textit{fixed}  and equal to $r$ across all auctions. We generate $M$ datasets with this specifications. By repeating out calculations on these $M$ datasets, we obtain confidence intervals for our empirical results. Throughout we set $n=100, \mu=1, \epsilon = 0.05, M=10$.
%

\paragraph{Real auction data} 
This dataset consists of the bids observed on Google's DoubleClick Ad Exchange for a random set of 100 buyers, 
submitted over three consecutive weekdays for a set of similar auctions.
We ensure that the items are similar by restricting the queries to a small industrialized upper-middle-income country and one type of device. 
For ease of interpretation, we scale the observed bids so they are in units of reserve price, 
i.e. on the same scale as the simulated auction data above with $r = 1$.
For each bidder we generate the missing/unobserved bids by resampling from the empirical distribution of her observed bids.

\begin{figure*}[h!]
       \centering
\subfigure{\includegraphics[width=0.32\textwidth]{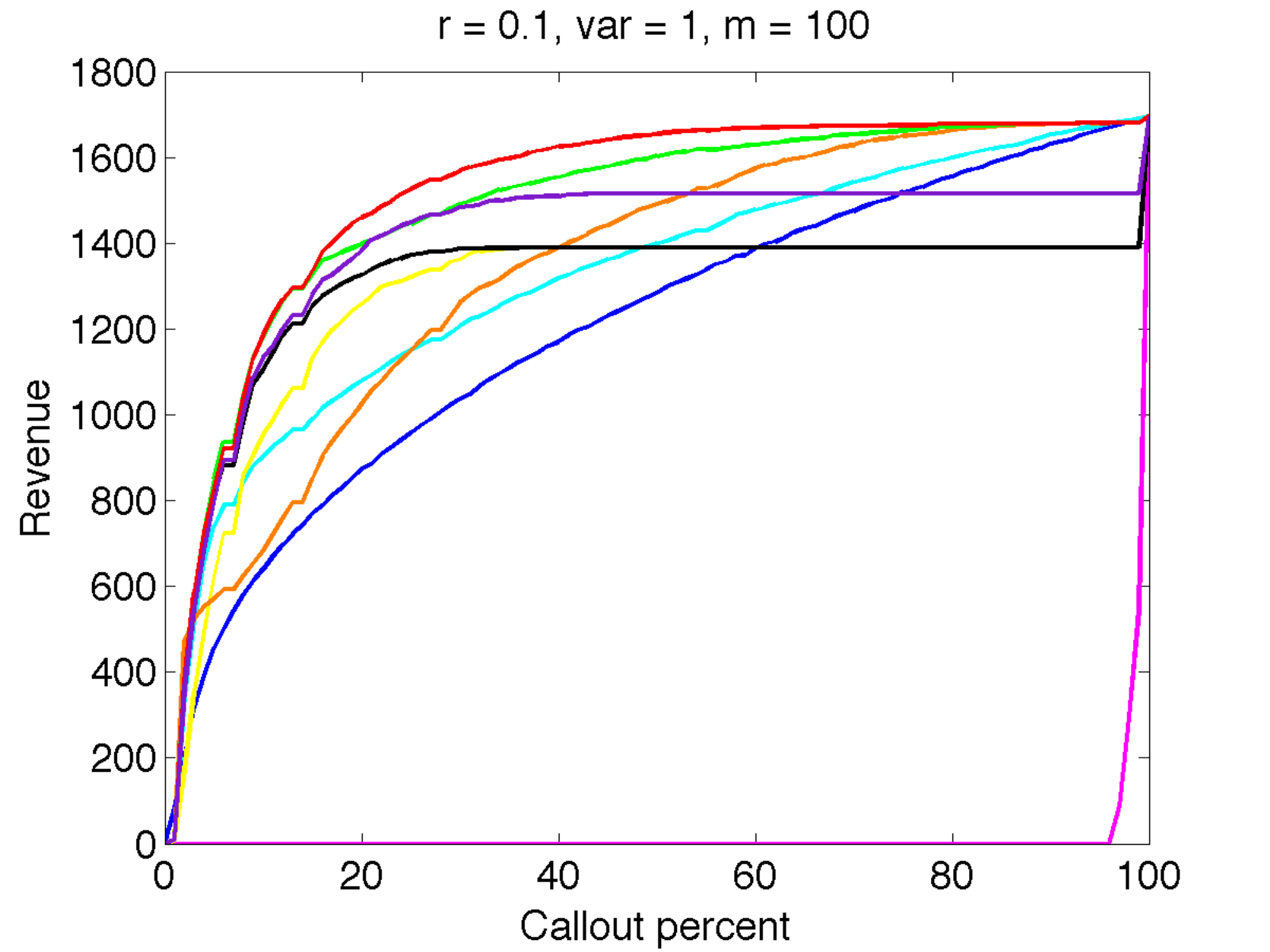}}
\subfigure{\includegraphics[width=0.32\textwidth]{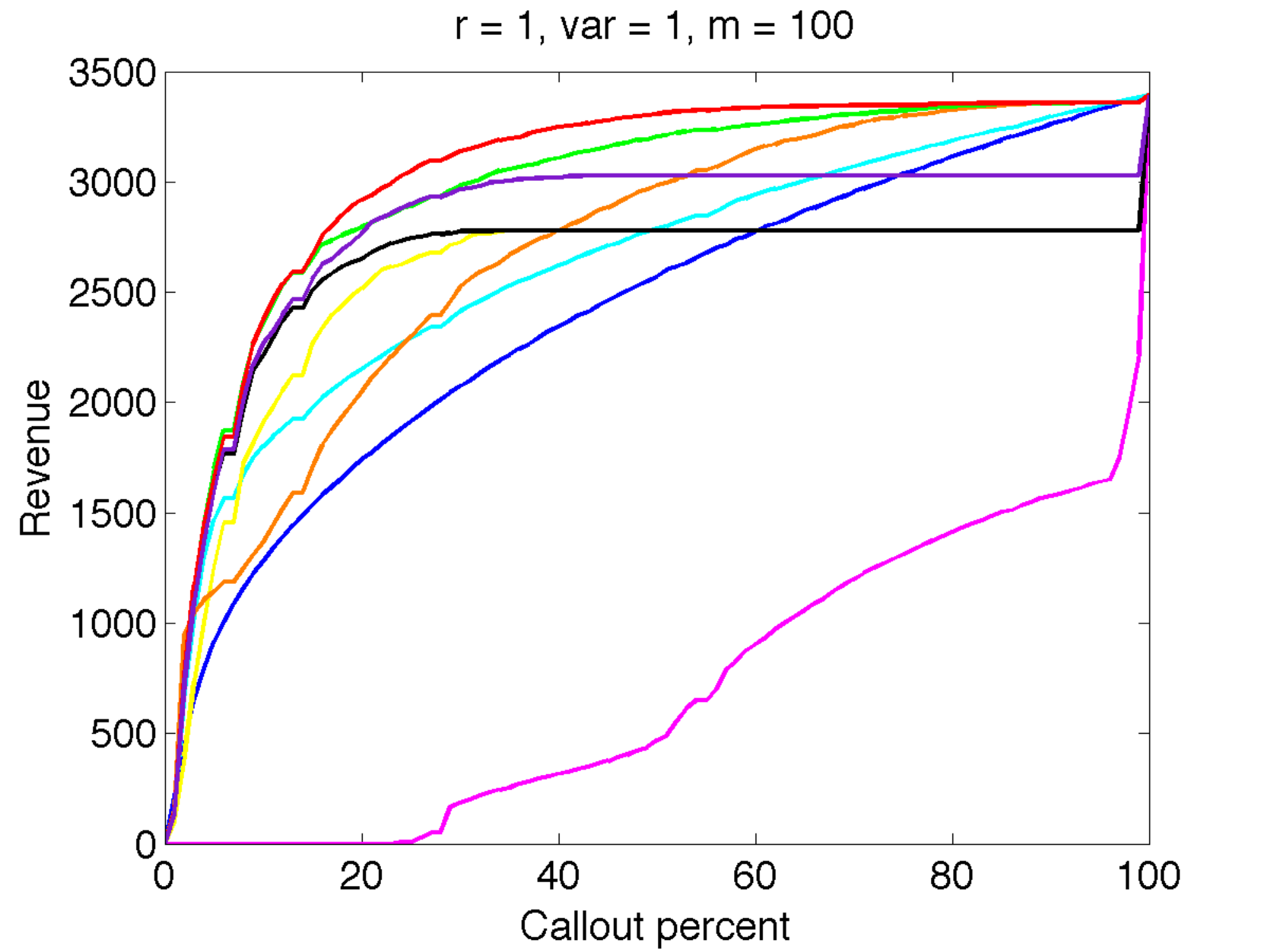}}
\subfigure{\includegraphics[width=0.32\textwidth]{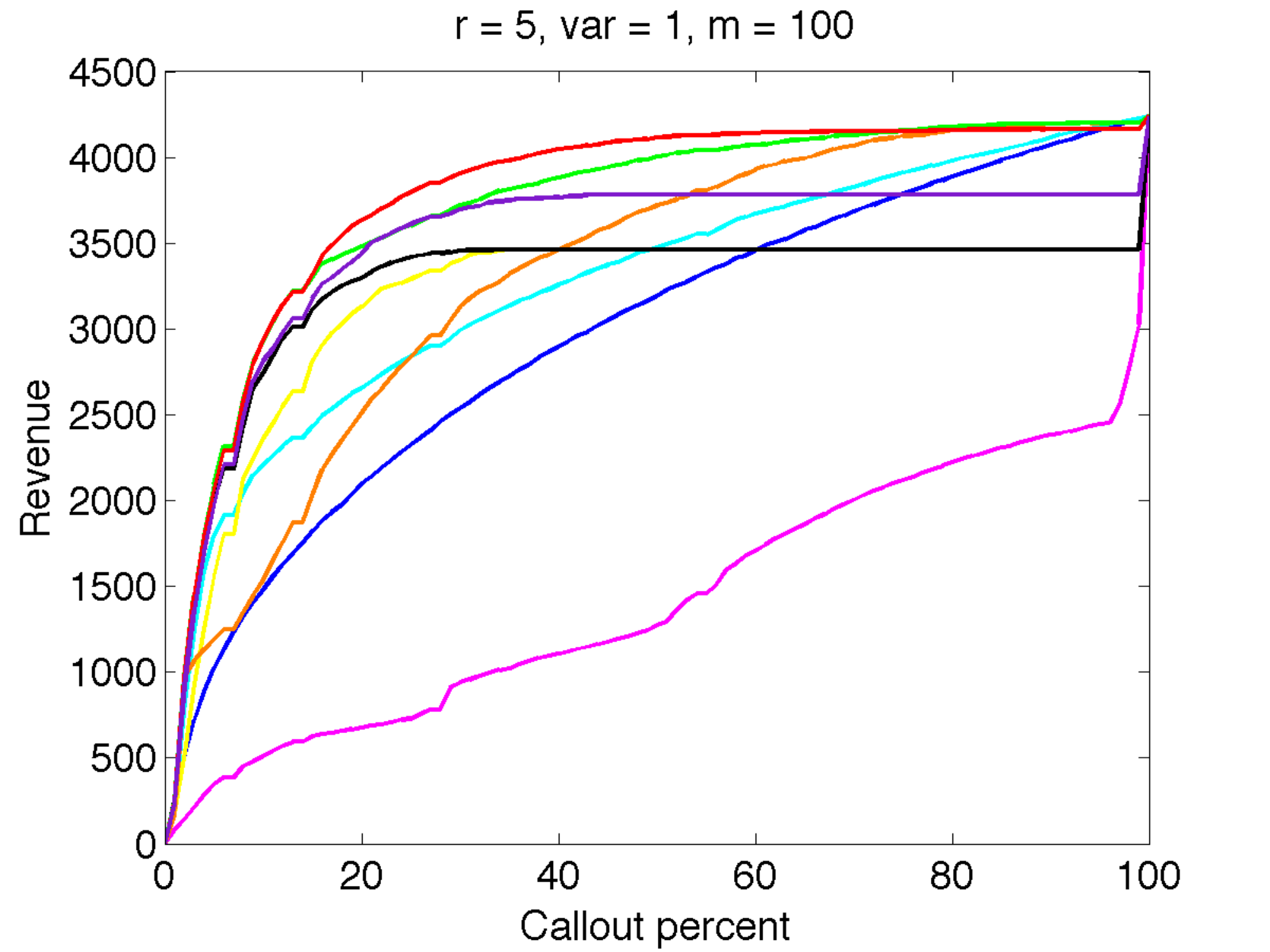}}
\subfigure{\includegraphics[width=0.32\textwidth]{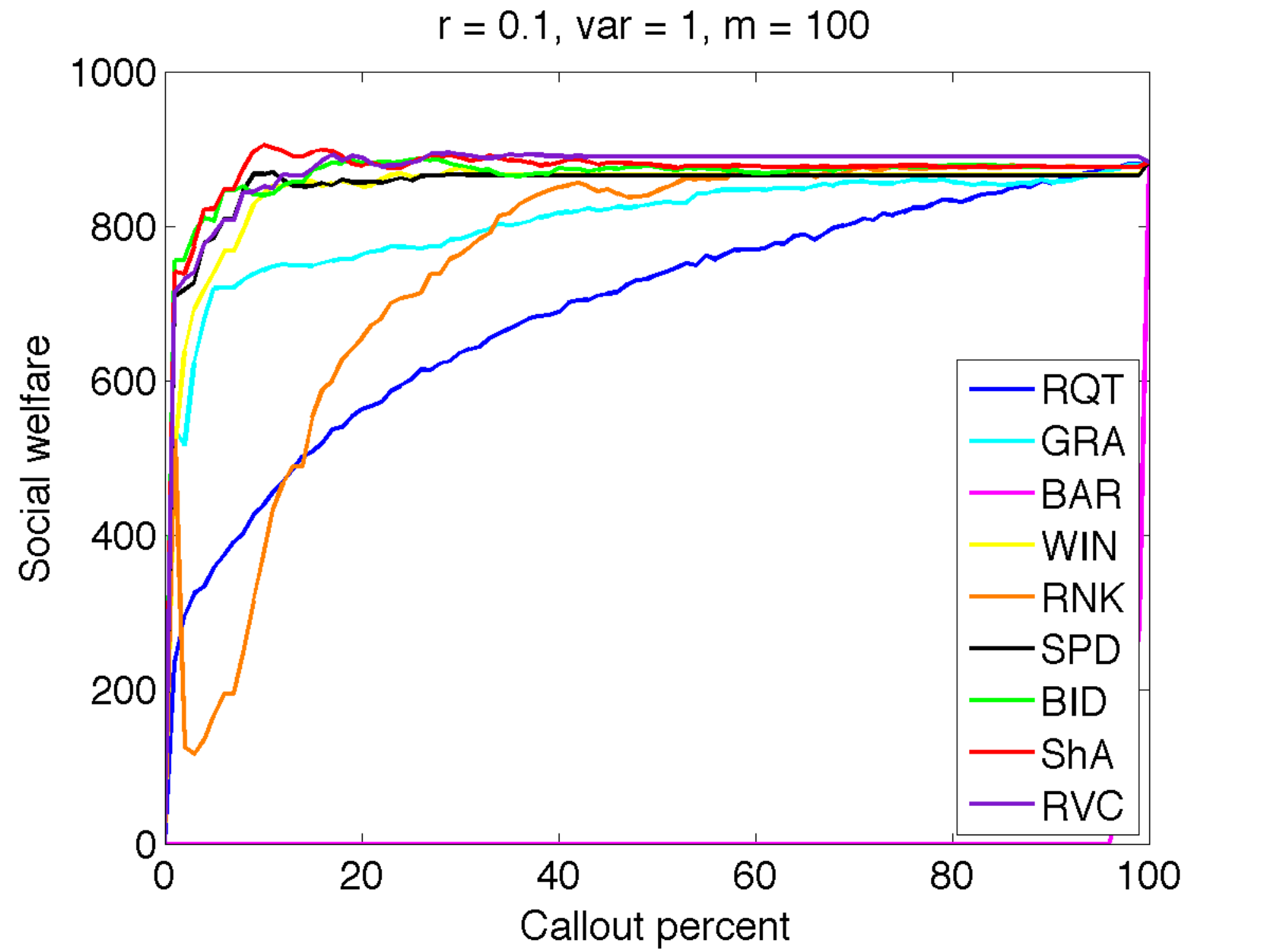}}
\subfigure{\includegraphics[width=0.32\textwidth]{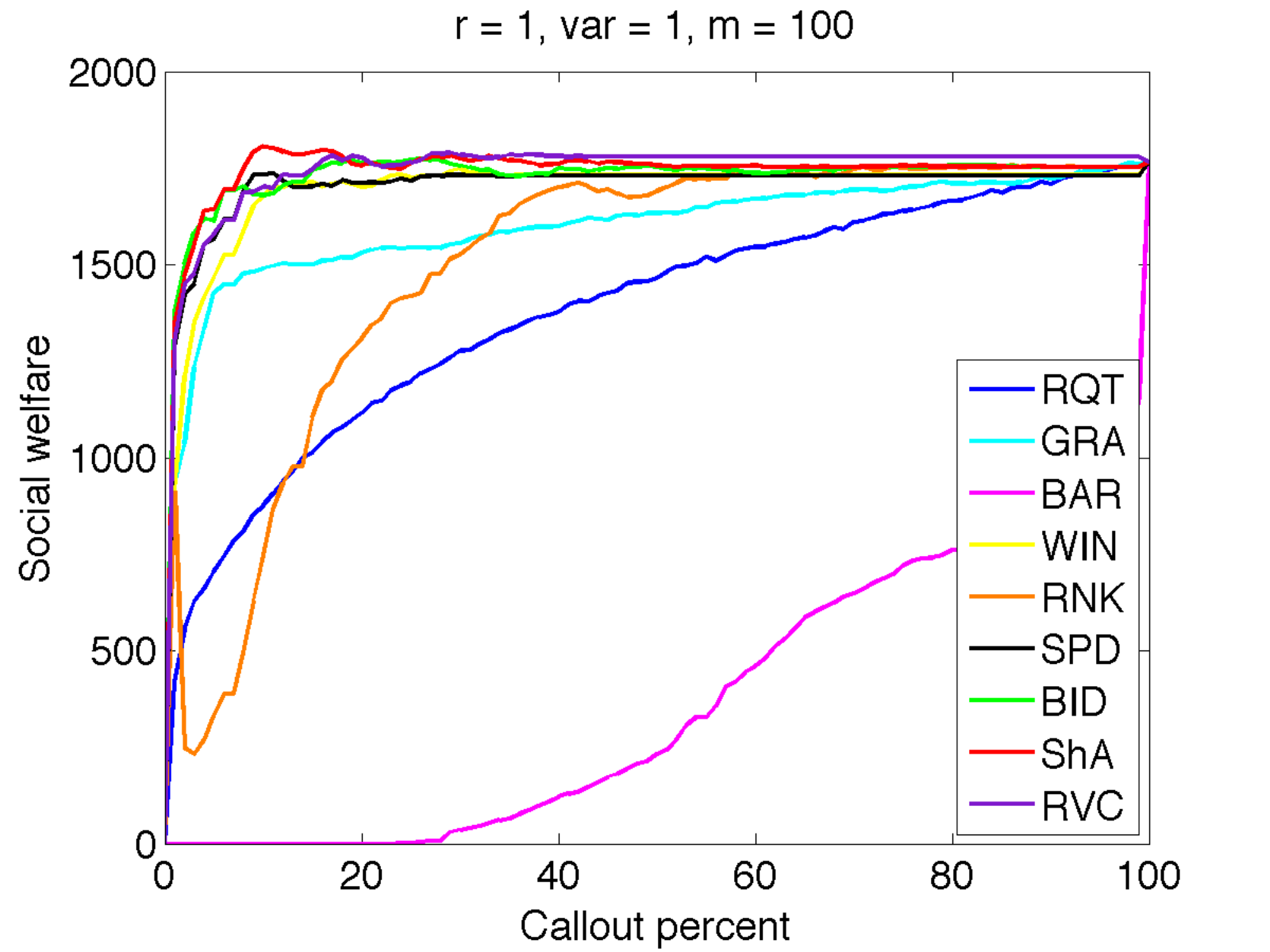}}
\subfigure{\includegraphics[width=0.32\textwidth]{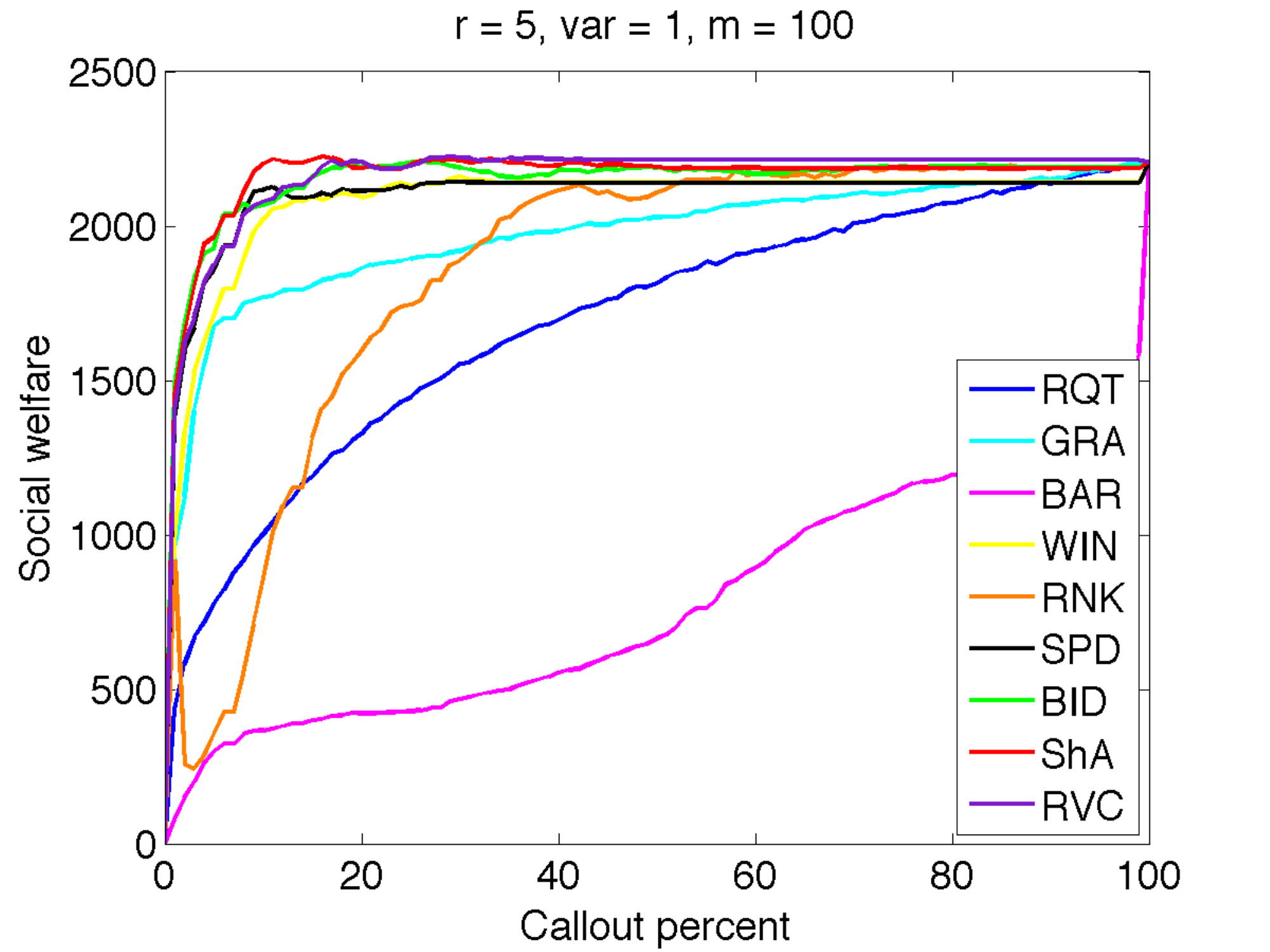}}
\subfigure{\includegraphics[width=0.32\textwidth]{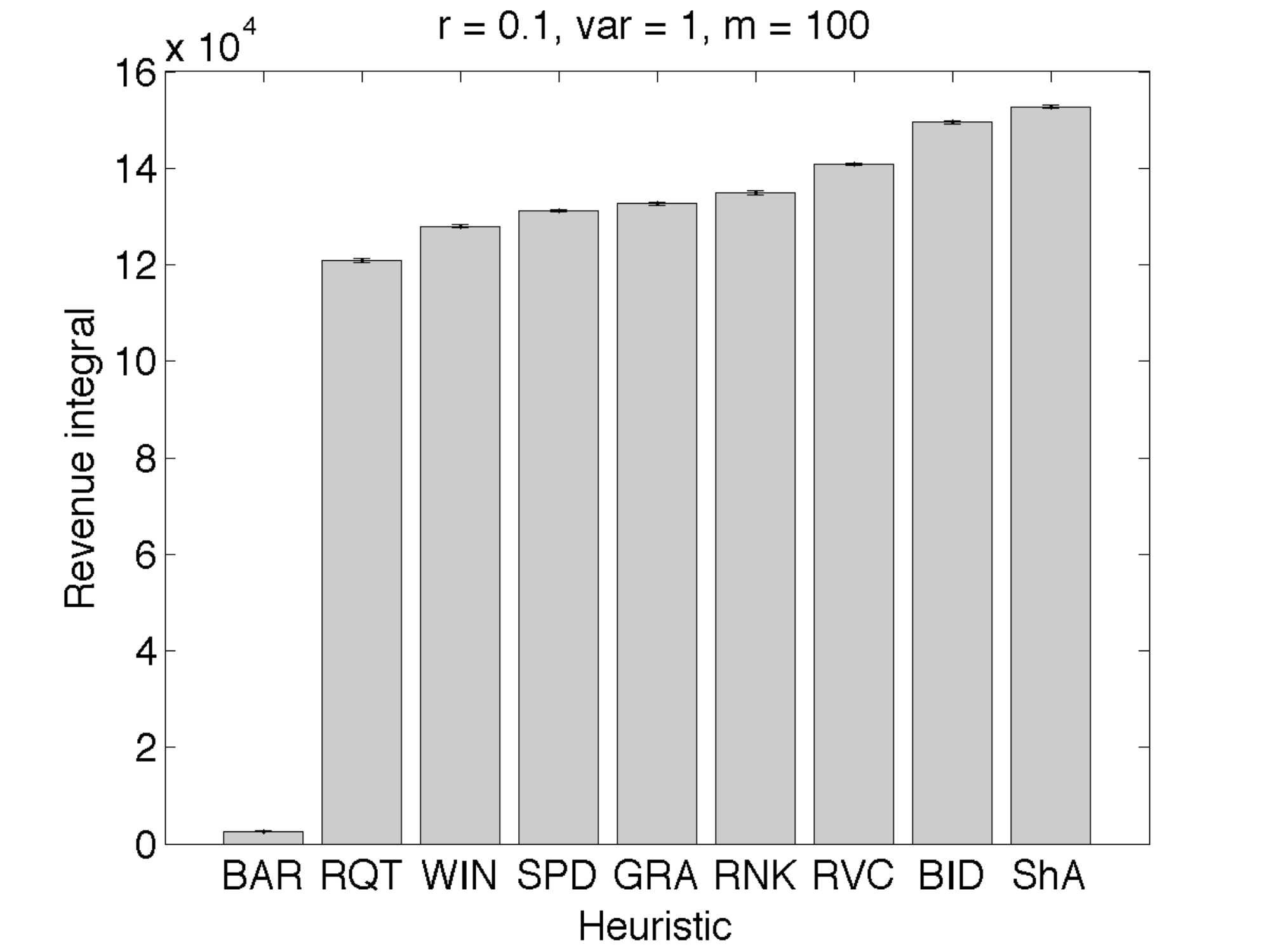}}
\subfigure{\includegraphics[width=0.32\textwidth]{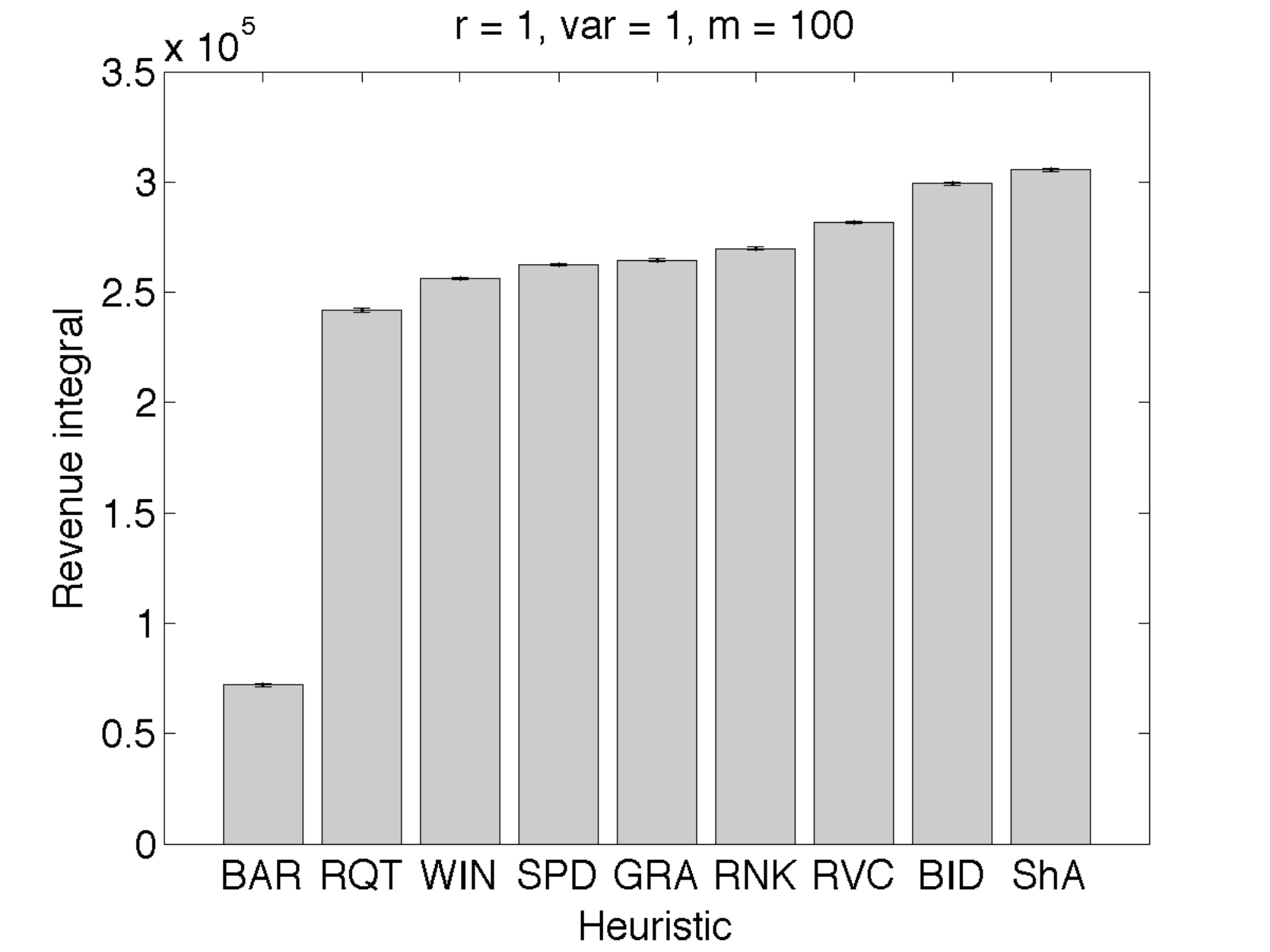}}
\subfigure{\includegraphics[width=0.32\textwidth]{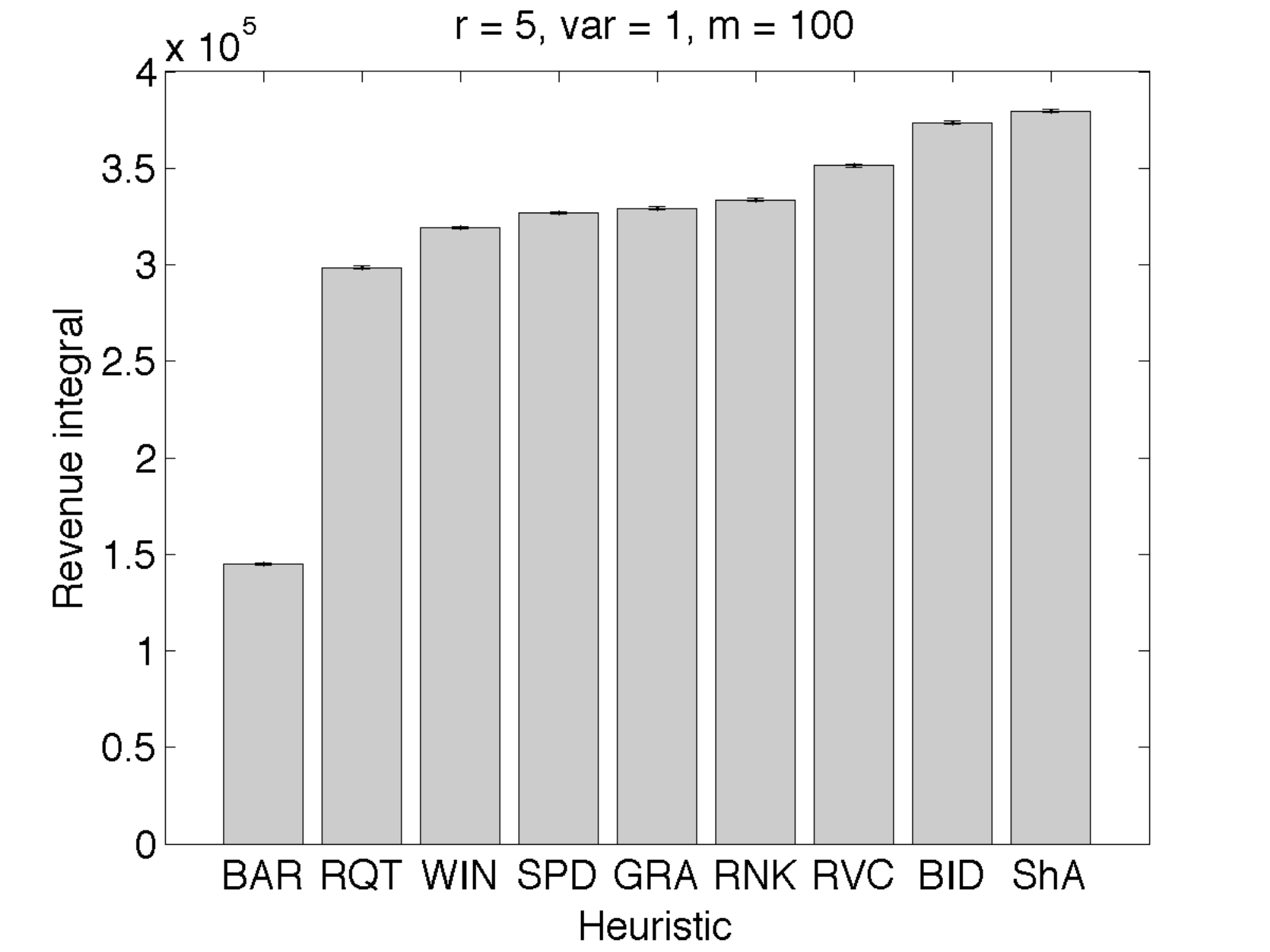}}

        \caption{The effect of reserve price on the performance of each callout mechanism. Here $T=100$ and $\sigma = 1$.} 
         \label{fig:reserve}
\end{figure*}

\begin{figure*}[t!]
       \centering
\subfigure{\includegraphics[width=0.32\textwidth]{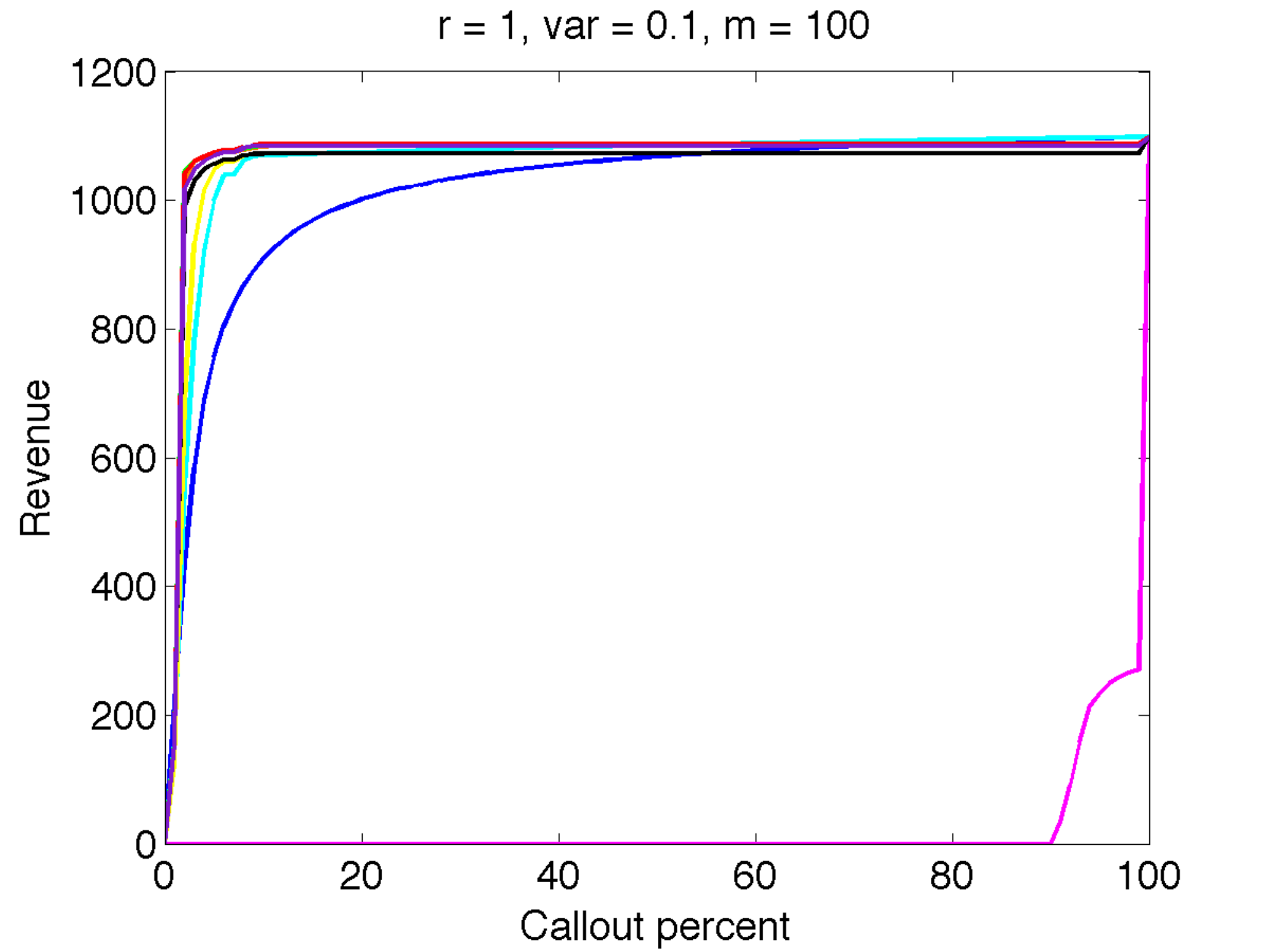}}
\subfigure{\includegraphics[width=0.32\textwidth]{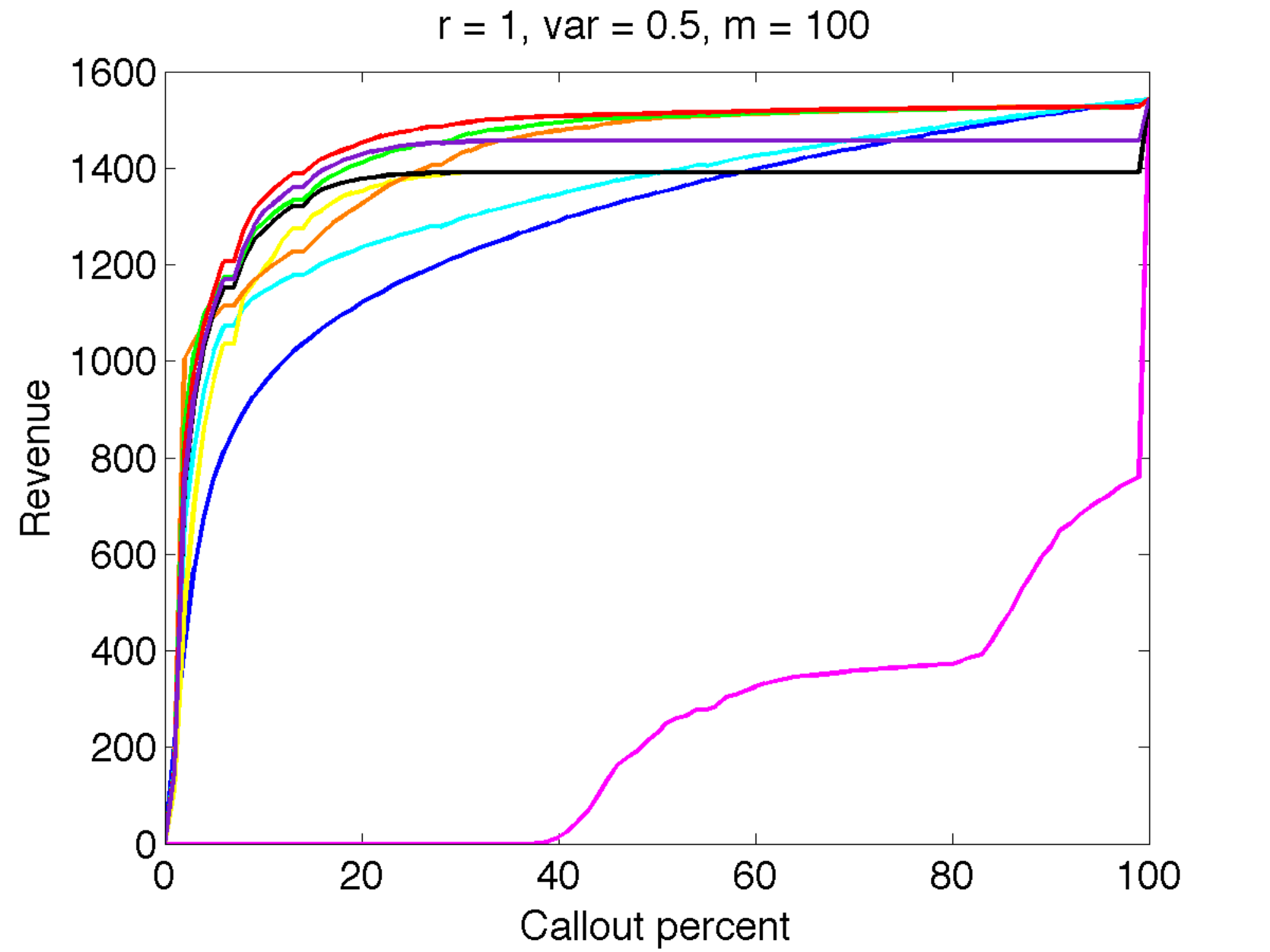}}
\subfigure{\includegraphics[width=0.32\textwidth]{figures/rv-n100m100r10mu1v10.pdf}}
\subfigure{\includegraphics[width=0.32\textwidth]{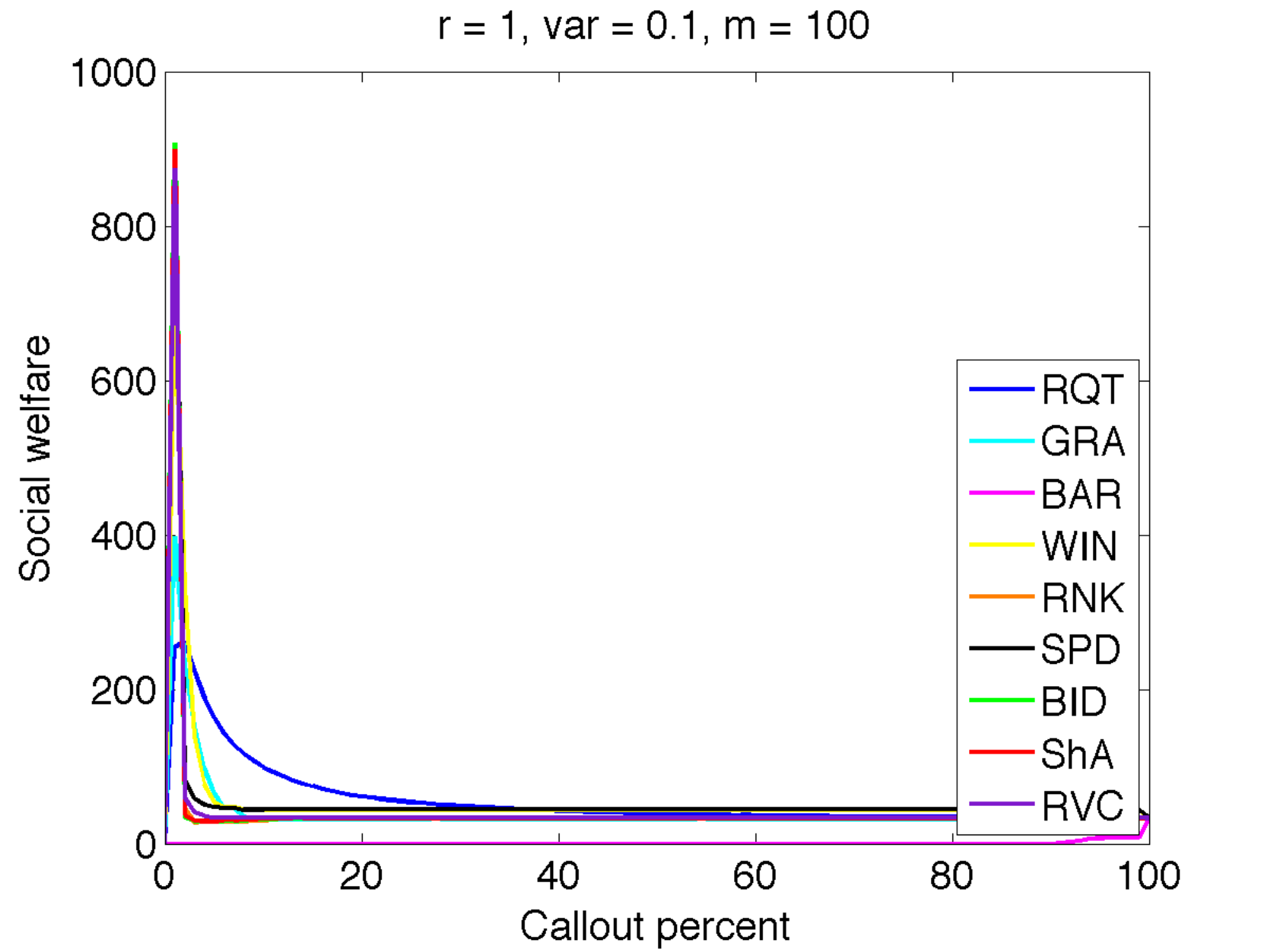}}
\subfigure{\includegraphics[width=0.32\textwidth]{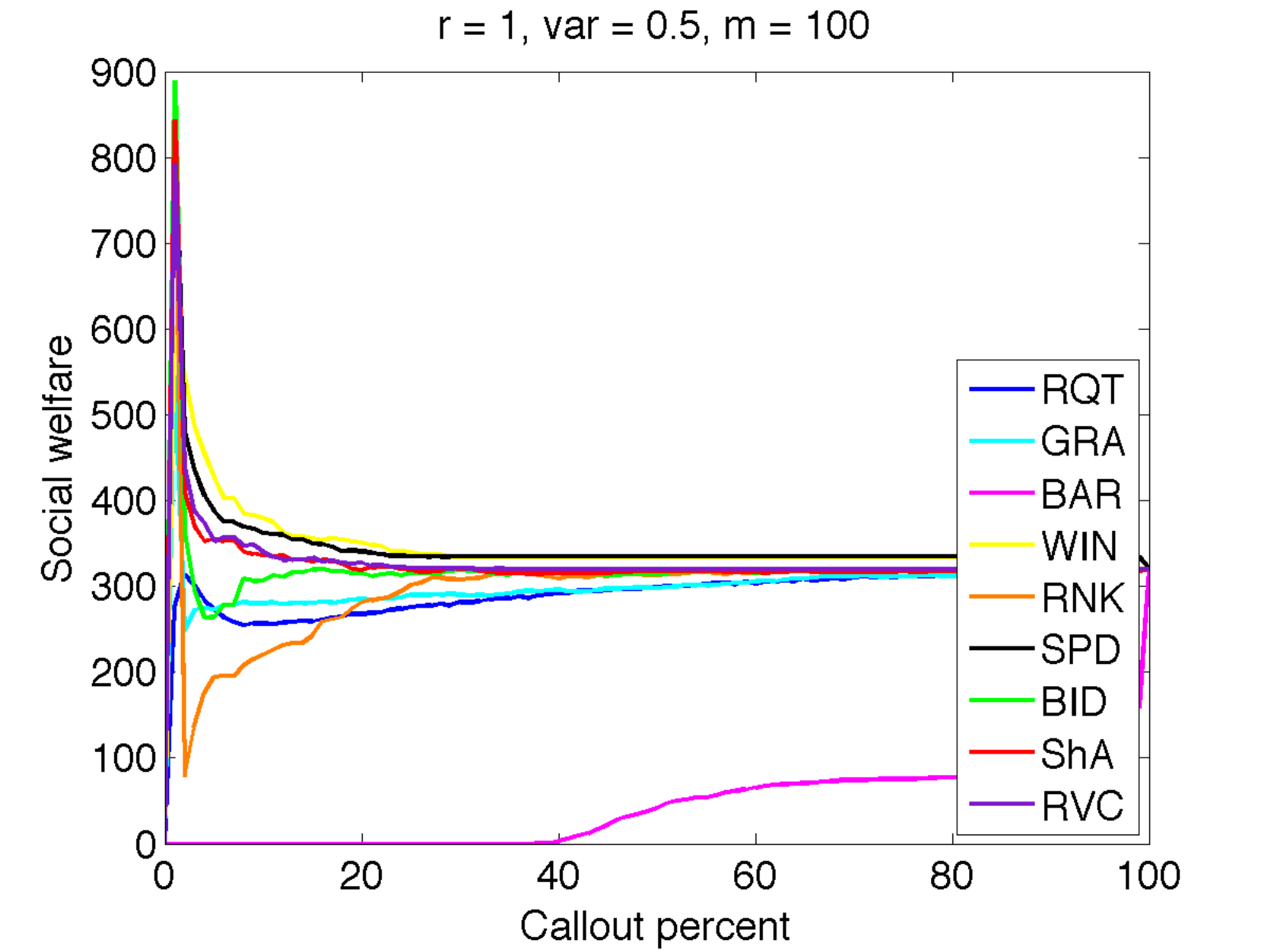}}
\subfigure{\includegraphics[width=0.32\textwidth]{figures/sw-n100m100r10mu1v10.pdf}}
\subfigure{\includegraphics[width=0.32\textwidth]{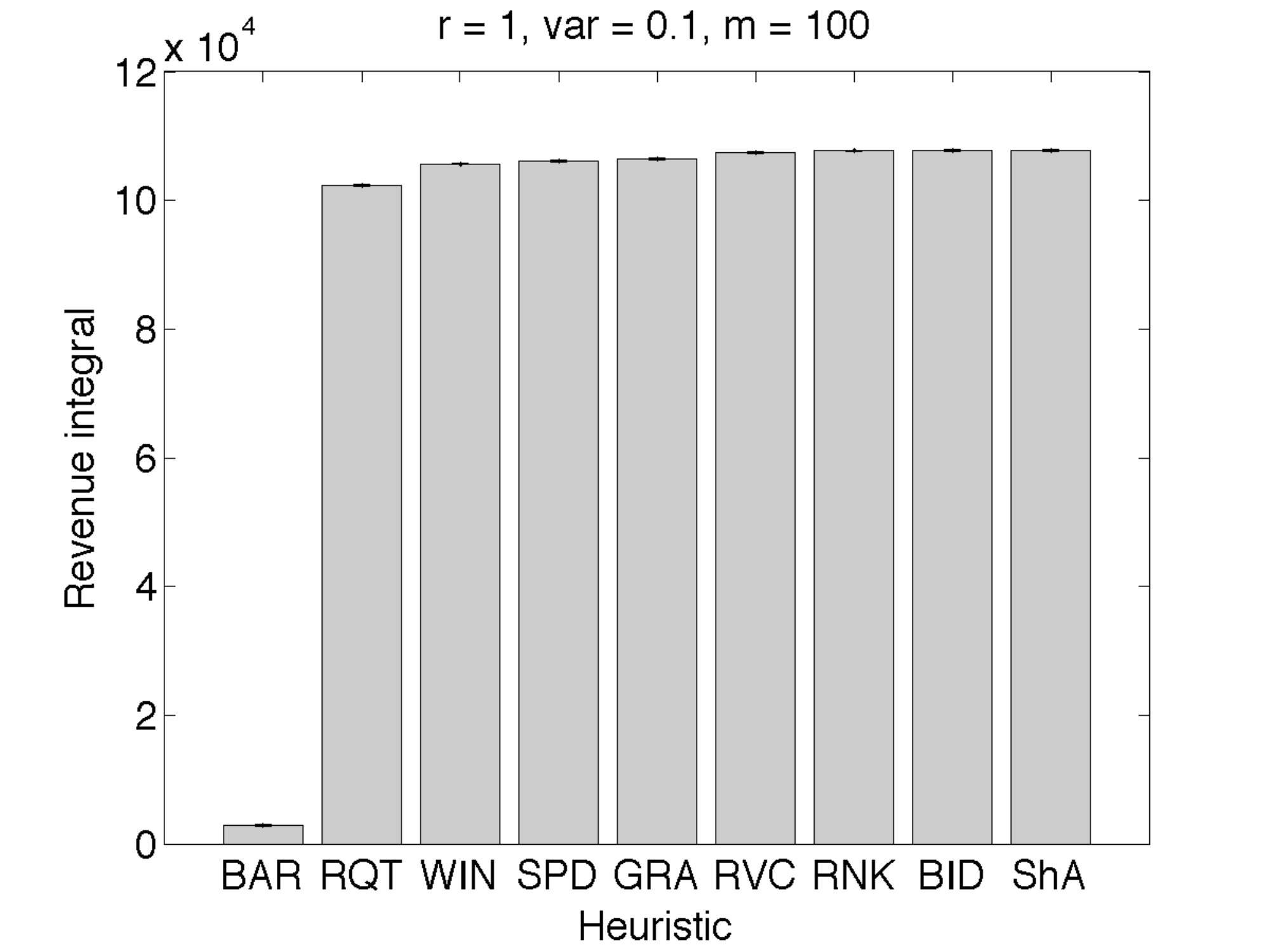}}
\subfigure{\includegraphics[width=0.32\textwidth]{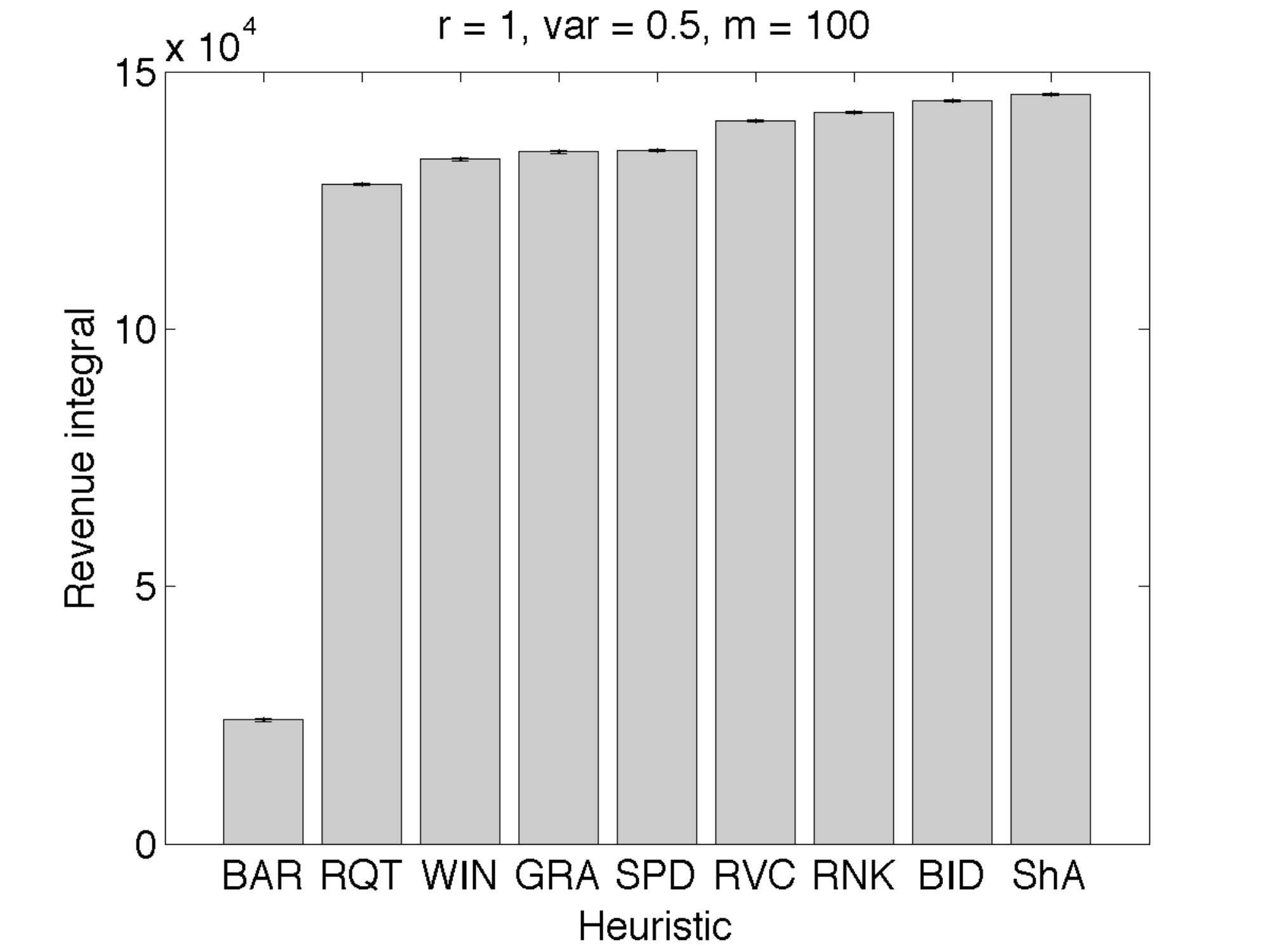}}
\subfigure{\includegraphics[width=0.32\textwidth]{figures/scores-n100m100r10mu1v10.pdf}}

        \caption{The effect of variance on the performance of each callout mechanism. Here $T=100$ and $r=1$.}
         \label{fig:variance}
\end{figure*}

\begin{figure*}[t!]
       \centering
\subfigure{\includegraphics[width=0.32\textwidth]{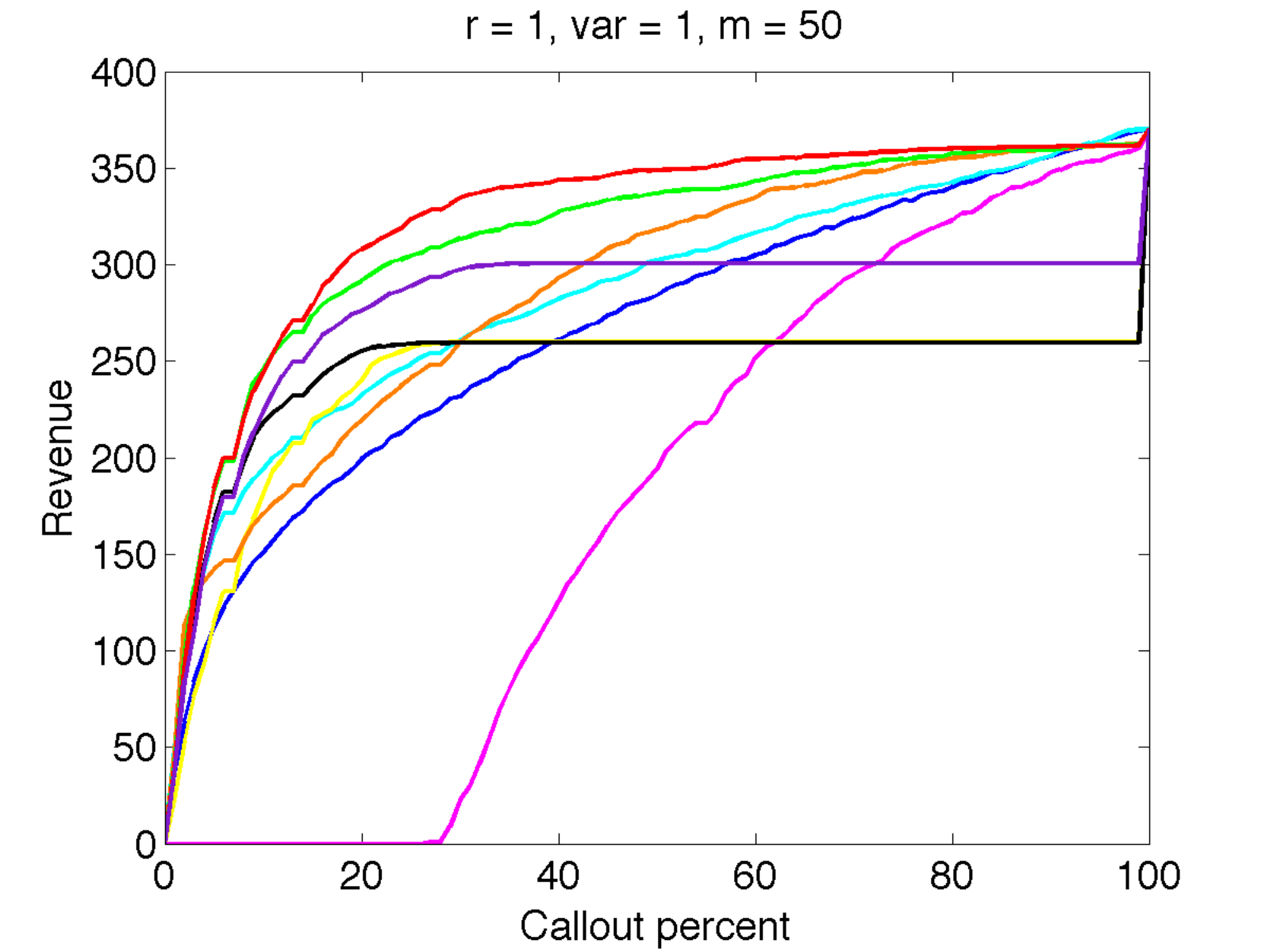}}
\subfigure{\includegraphics[width=0.32\textwidth]{figures/rv-n100m100r10mu1v10.pdf}}
\subfigure{\includegraphics[width=0.32\textwidth]{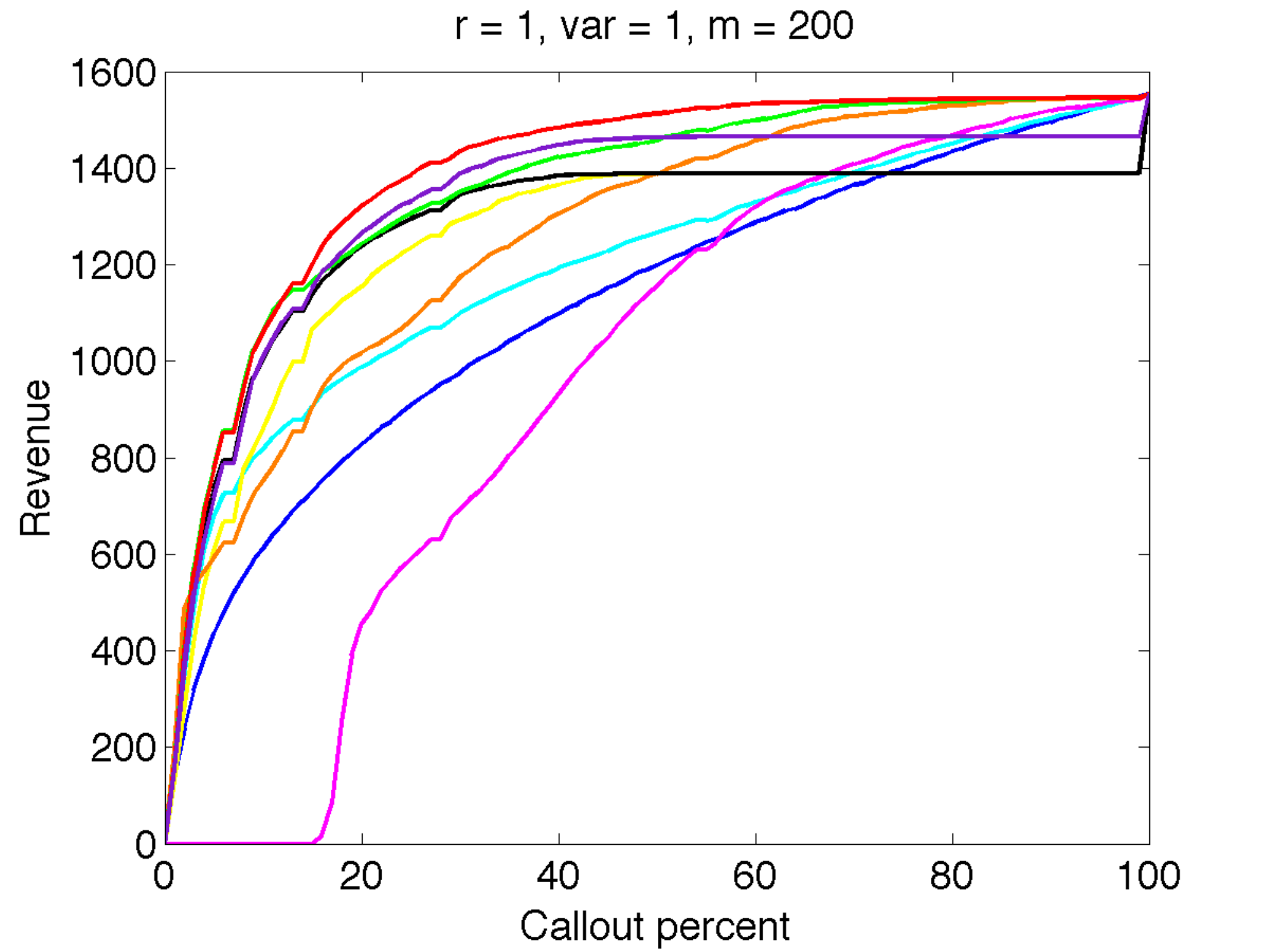}}
\subfigure{\includegraphics[width=0.32\textwidth]{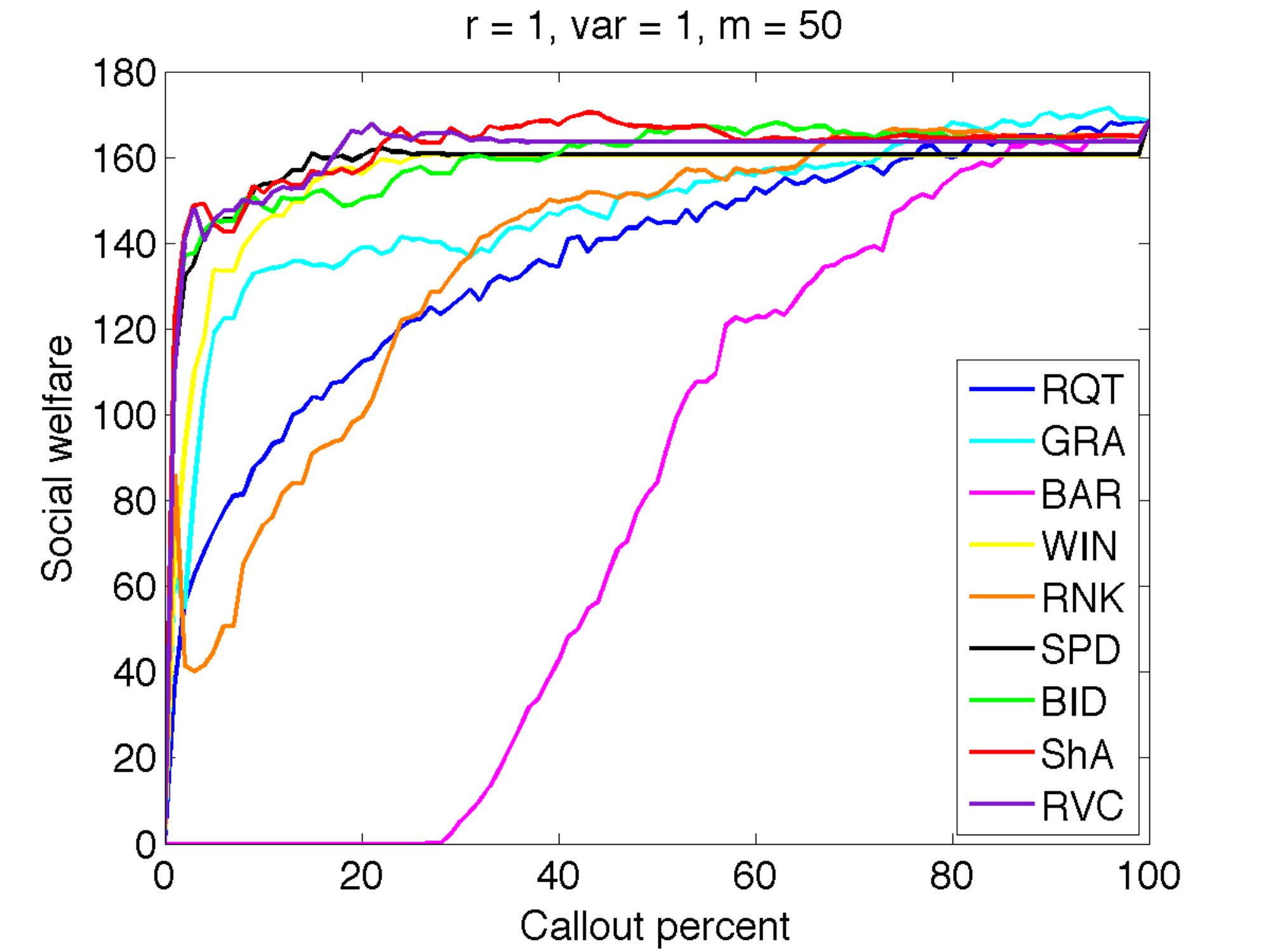}}
\subfigure{\includegraphics[width=0.32\textwidth]{figures/sw-n100m100r10mu1v10.pdf}}
\subfigure{\includegraphics[width=0.32\textwidth]{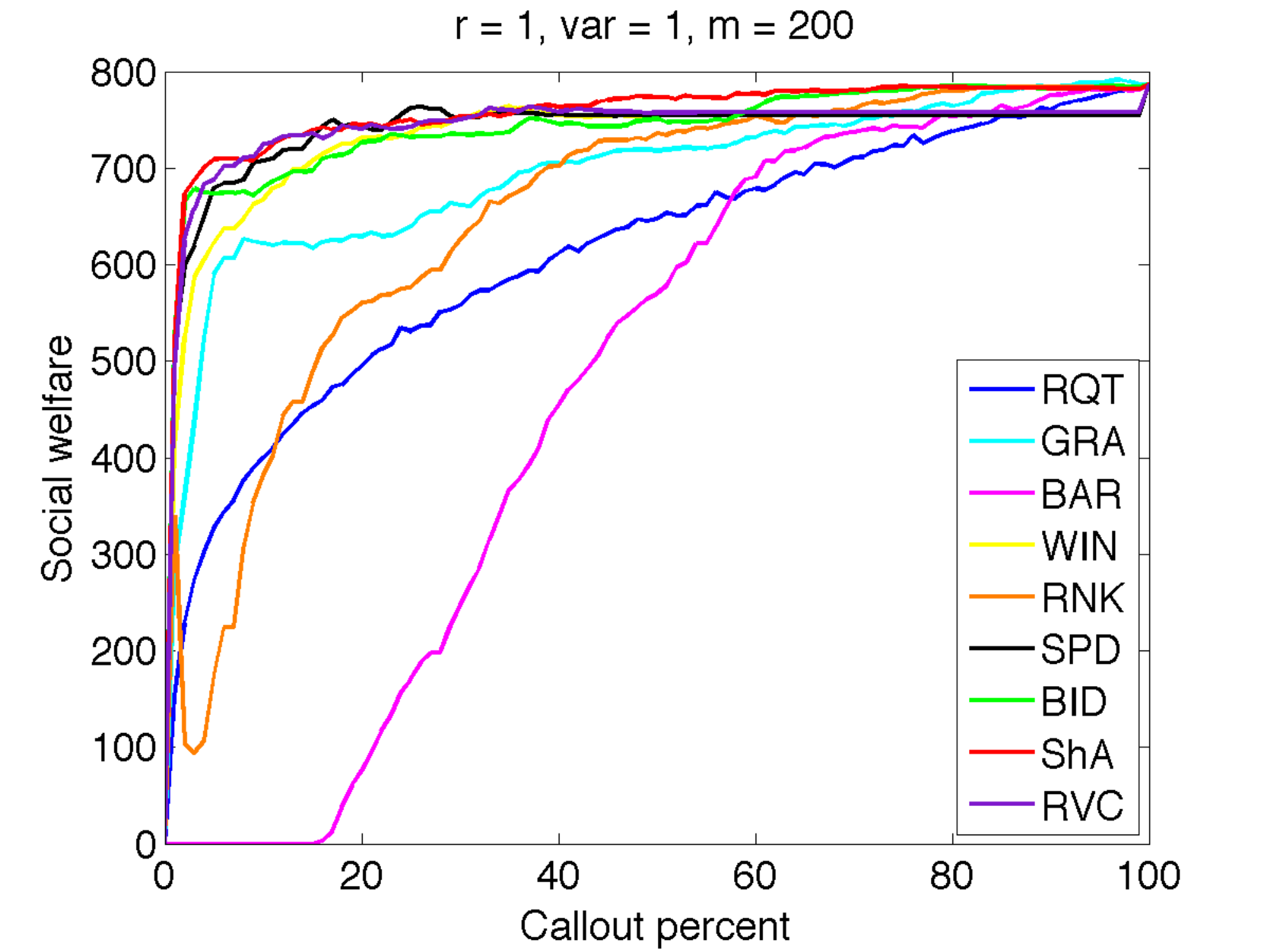}}
\subfigure{\includegraphics[width=0.32\textwidth]{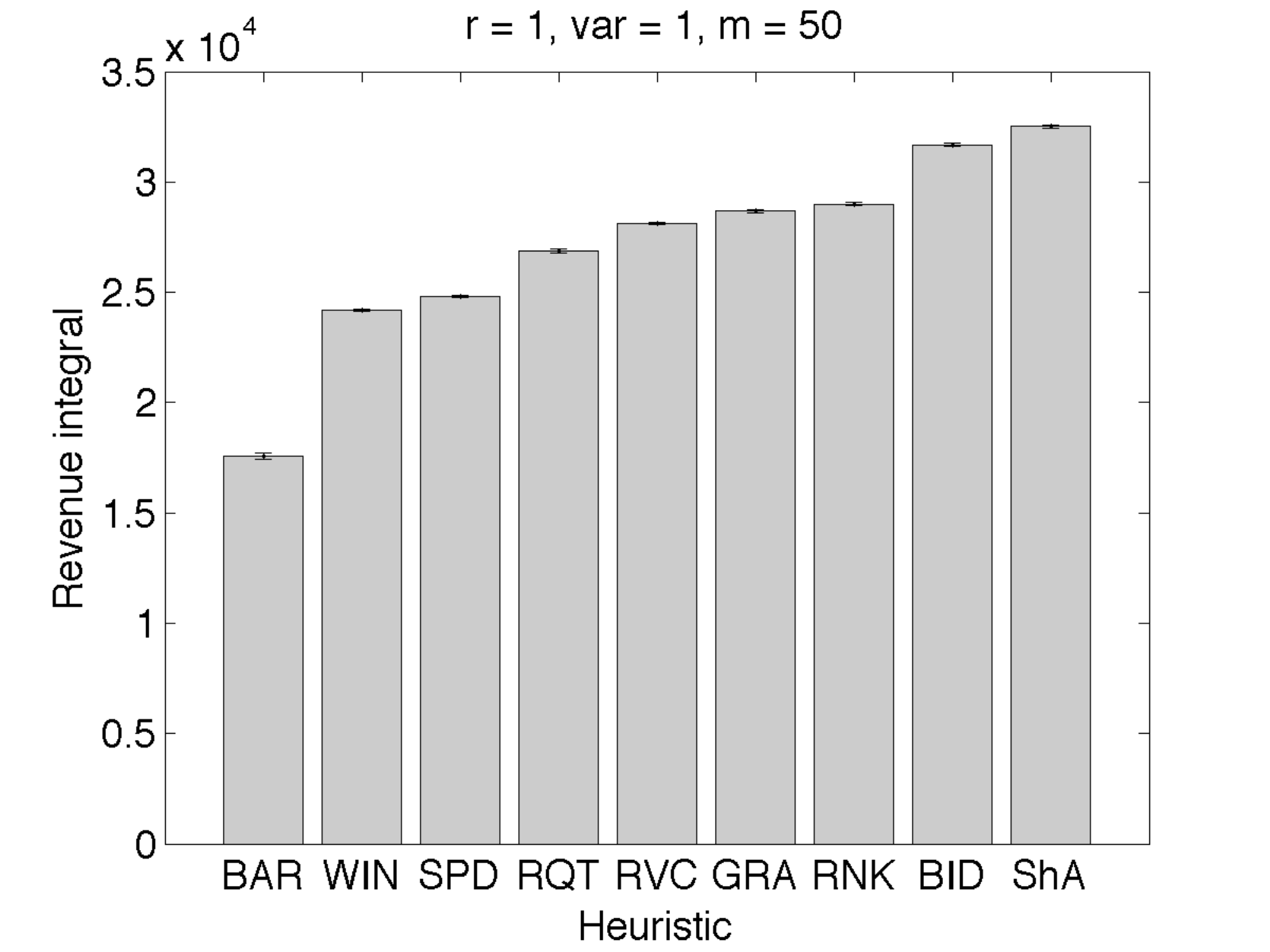}}
\subfigure{\includegraphics[width=0.32\textwidth]{figures/scores-n100m100r10mu1v10.pdf}}
\subfigure{\includegraphics[width=0.32\textwidth]{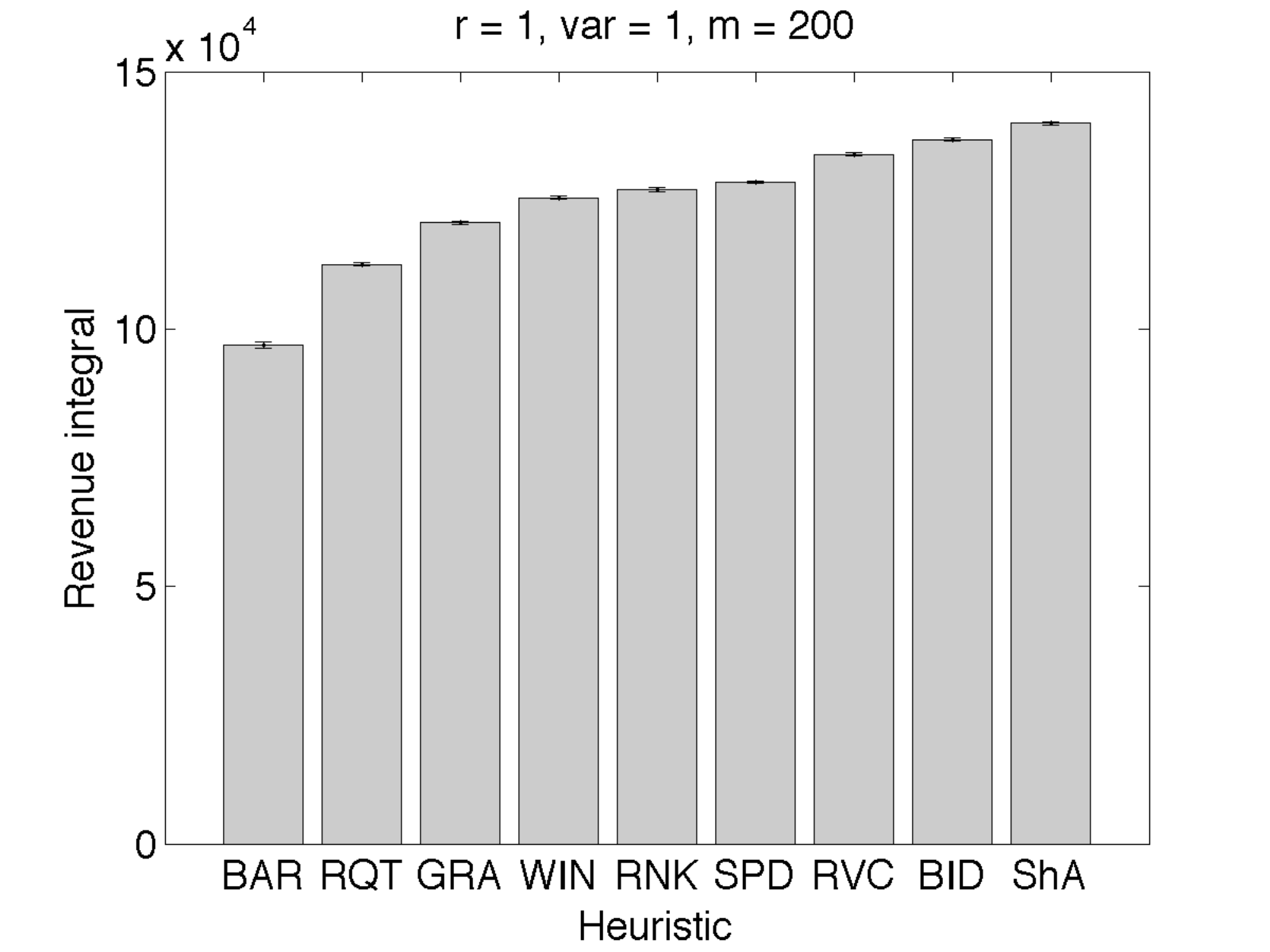}}

        \caption{The effect of $T$ on the performance of each callout mechanism. Here $r=1$ and $\sigma= 1$.}
         \label{fig:m}
\end{figure*}

\subsection{Findings}
\begin{figure*}[t!]
       \centering
\subfigure{\includegraphics[width=0.32\textwidth]{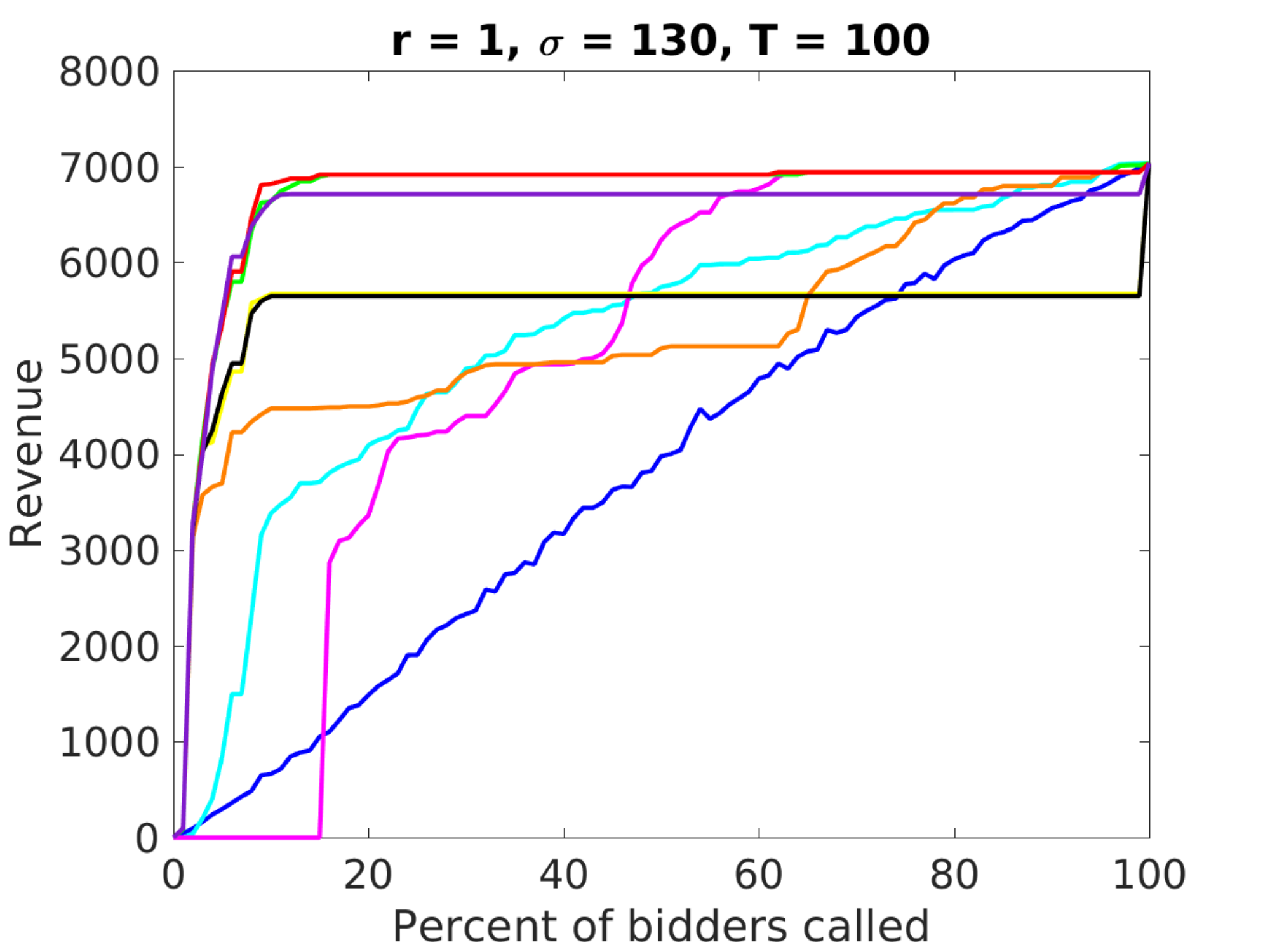}}
\subfigure{\includegraphics[width=0.32\textwidth]{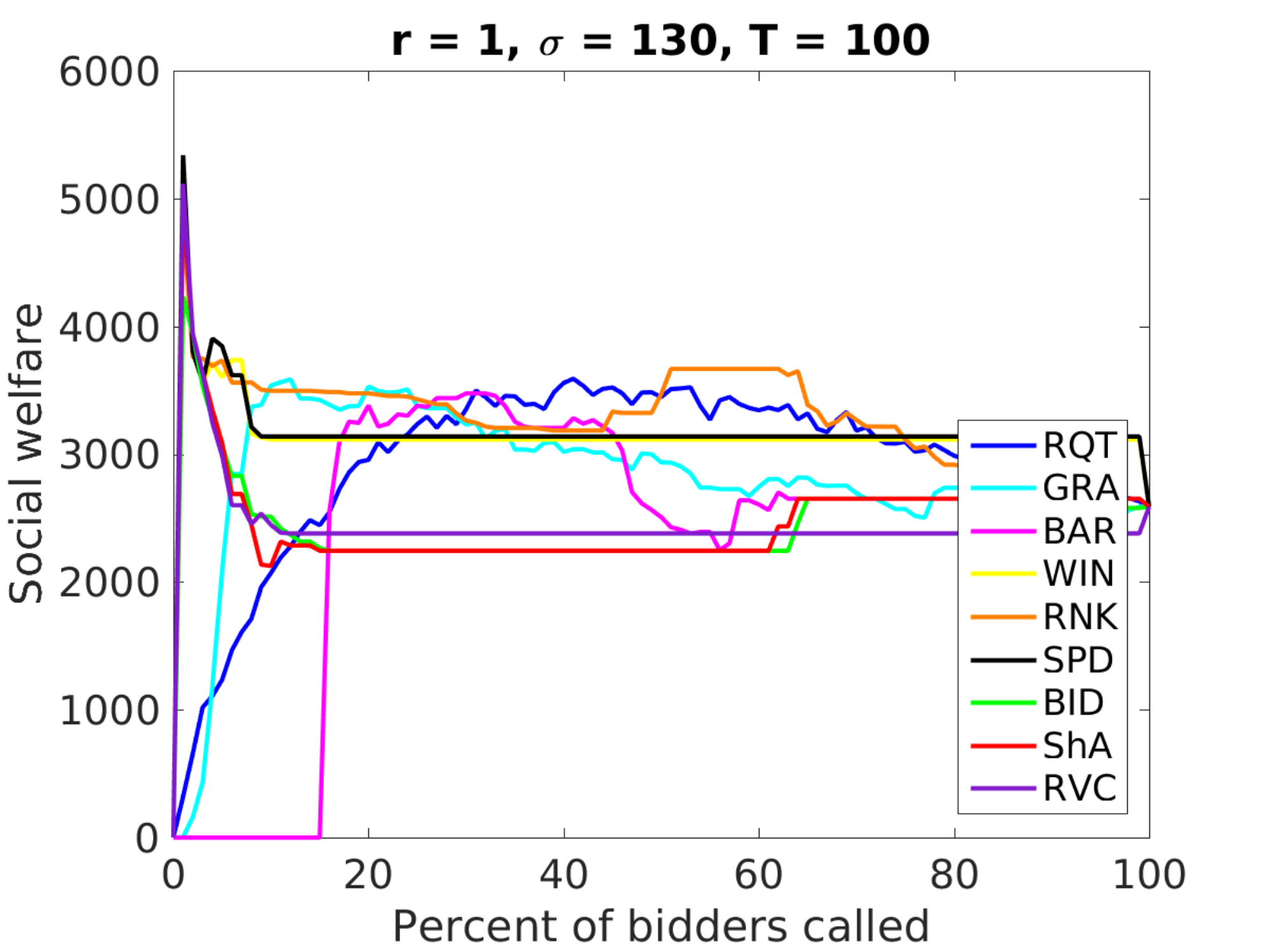}}
\subfigure{\includegraphics[width=0.32\textwidth]{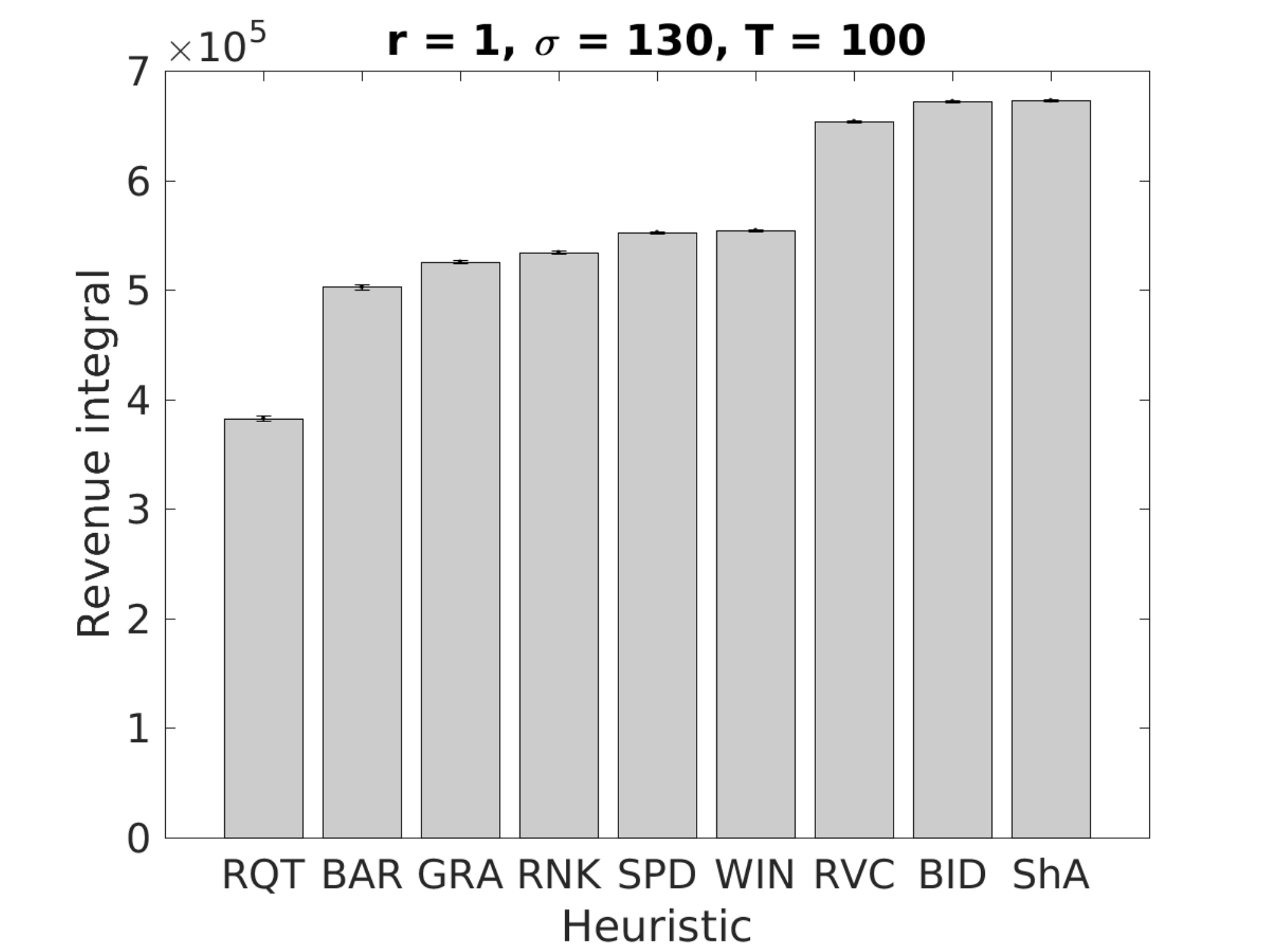}}
        \caption{The performance of each callout mechanism on the real auction dataset. Here $n=100$, $\sigma= 130$, and $r=1$.}
         \label{fig:real}
\end{figure*}

Figures \ref{fig:reserve}, \ref{fig:variance}, \ref{fig:m} illustrate the performance of our callout mechanisms along with the baselines on the synthetic dataset for different setting of the parameters. The first row in each figure depicts the revenue and the second row depicts the social welfare versus the percentage of bidders called out to each auction. The third row depicts the average revenue earned by each heuristic across all threshold values. This can be thought of as an overall score for each algorithm. Figure~\ref{fig:real} illustrates the performance of our callout mechanisms on the real dataset.

Figure \ref{fig:reserve} illustrates the effect of the reserve price $r$ on the performance of each callout mechanism when the average variance $\sigma$ across bidders is $1$ and the number of items to be auctioned off, $T$, is $100$.  We observe that regardless of the reserve price, the relative rank of the heuristics in terms of the revenue integral, remains unchanged. For example ShA always outperforms the other algorithms by an statistically significant margin. Also see Figure~\ref{fig:real} for a similar trend. As we expect, the performance of BAR improves as $r$ grows larger. 
Note that when the percentage of bidders called to each auction is high, WIN, SPD and RVC fail to beat the RQT baseline. The reason is that these metrics give high scores to only a small subset of bidders and assign the remainder of bidders scores that are about equal. This induces the flat interval in the corresponding curves.
In terms of social welfare, our heuristics always outperform RQT, except for BAR and RNK. RNK does not maintain sufficient social welfare when the percentage of bidders called to the auction is low. The reason is obvious: RNK calls bidders with lowest (best) ranks, making the winner pay more and degrading the social welfare as the result.

Figure \ref{fig:variance} illustrates the effect of average bidding variance $\sigma$ on the performance of each callout mechanism when the reserve price $r$ is fixed and equal to $1$ and the number of items $T$ is $100$. 
As variance increases, we see a divergence in the performance of various heuristics. In particular, the performances of GRA, BAR, RNK, and BID all start to deteriorate.
Also, when the percentage of called bidders is small, we observe a sudden jump in social welfare. The heuristics exhibiting this phenomenon fail to maintain the auction pressure, dropping the winner's close competitors. As the result, the winner pays less, which boosts the social welfare.

Figure \ref{fig:m} illustrates the effect of the number of items $T$ on the performance of each callout mechanism when the reserve price $r$ is fixed and equal to $1$ and the average variance $\sigma$ across bidders is $1$. 
We see that as $T$ increases, the difference between the performance of different heuristics begins to vanish: the performance of all algorithms (expect RQT) converge to that of ShA, even WIN. 

The main takeaway messages from the empirical results presented above are:
\begin{itemize}
\item \textit{It is easy to beat the RQT baseline.} Even our crudest heuristics, WIN and SPD, outperform RQT most of the time. 
\item \textit{Some of our heuristics outperform both baselines.} More sophisticated heuristics, e.g. RVC, RNK, BID, and ShA, consistently outperform the baselines.
\item \textit{A good callout mechanism can significantly improve revenue. } For example in certain settings, ShA results in revenue up to 50\%  more than that of RQT and 25\% more than that of GRA.
\item \textit{ShA $>$ BID$>$ \{ RNK, RVC, GRA\} $>$ \{SPD, WIN, RQT, BAR \}.} And these results are statistically significant across the settings investigated here. 
\item \textit{The more information a heuristic contains about the revenue impact of bidders, the better it performs.} We believe this is why ShA outperforms all the other heuristics: As we noted earlier, ShA estimates the counterfactual revenue impact of each bidder, and as a result improves revenue the most. Overall better heuristics call out more specifically to bidders with greater revenue impact.
\end{itemize}

%% file: conclusion.tex
\section{Conclusion and Future Directions}\label{sec:conclusion}
We presented a framework for evaluating the performance of callout mechanisms in repeated auctions using historical data only. Our framework is general enough to enable the study of other heuristics in settings beyond those considered here (e.g. alternative auction mechanisms, bidding distributions, etc.).
In future, we intend to investigate the performance of more complicated callout mechanisms, including ones with more sophisticated learning steps; 
ones that combine multiple heuristics in a single score; ones that target bidders by means not easily represented by single-metric thresholding; and mechanisms that use online dynamic (as opposed to myopic) targeting.


%% file: technical.tex
\section{Omitted Technical Details}\label{app:tech}

\subsection{Proof of Proposition~\ref{prop:hardness}}
We show that this class of problems includes a known NP-hard problem---maximum 3-dimensional matching~\cite{3DimMatching}---as a special case. Consider the following setting: Let $\delta=1$. Given an instance of the 3-dimensional matching, construct an instance of our problem as follows: consider an item $i_y$ for each $y \in Y$ and a bidder $b_x$ for each $x \in X \cup Z$.  All bidders have capacity 1 (i.e. $q_x = 1$). If there is an edge between $x \in X \cup Z$ and $y \in Y$ in the 3-dimensional matching problem, $b_x$ bids 1 for $i_y$, otherwise she bids $0$ (that is, the bidding distributions are all degenerate with all the mass on either 0 or 1). Let the reserve price $r_y$ be 0 for all $y \in Y$. This means that to sell item $i_y$ for a price greater than zero we need to call at least two bidders to the auction for $i_y$. In this case the winner will pay exactly $\$ 1$ for the item. Also note that calling more than two people to the auction cannot increase the auction's revenue. 
With this reduction it is easy to see that the maximum 3-dimensional matching in the original problem is precisely the callout that maximizes the exchange's revenue and vice versa. \qed

\subsection{Examples}\label{app:ex}
\begin{figure}[h!] 
\centering 
\includegraphics[width=0.7\textwidth]{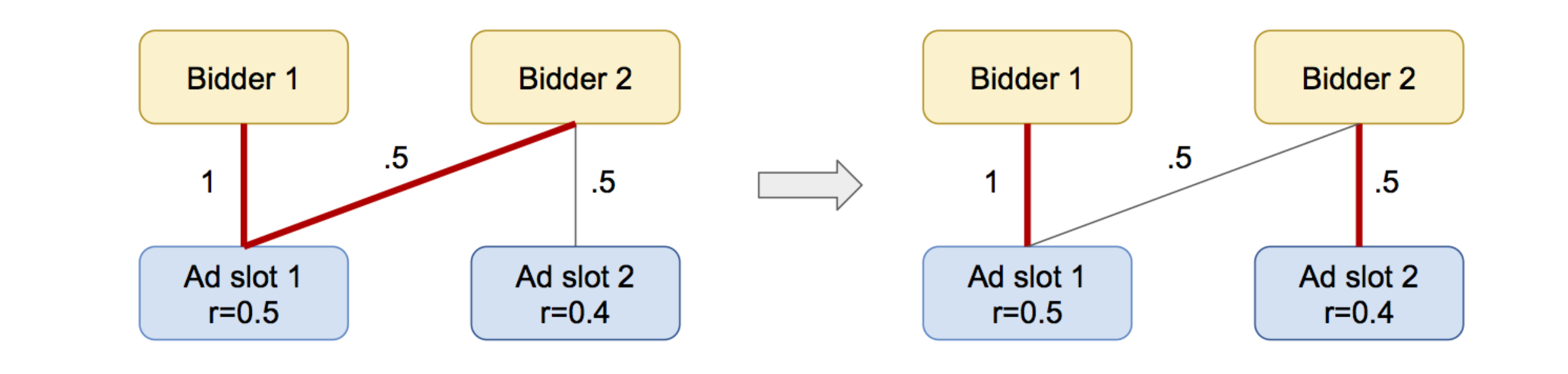}
\caption{Illustration of Example \ref{ex:1}. Callout mechanism (a) on the left and (b) on the right.} 
\label{fig:ex1}
\end{figure}

\begin{example}\label{ex:1}
Consider the following setting (See Figure~\ref{fig:ex1}): Suppose $A =\{a_1, a_2 \}$, $B = \{1,2\}$, $q_1= q_2 = 1$, and $\delta=1$.
Assume $b^1_{1} = 1, b^2_{1}=0$ and $b^1_{2} = b^2_{2}=0.5$. Also $r_1= 0.5$ and $r_2= 0.4$. The callout mechanism (a) assigns bidder 1 to $a_1$ and randomly decides which auction to call the second bidder to. This results in an expected revenue to the exchange equal to $0.5 + 0.5 \times 0.4=0.7$ and utility for the bidders equal to $0.5$ and $0.05$, respectively. The callout mechanism (b) assigns bidder 1 to $a_1$ and bidder 2 to $a_2$ and results in exchange revenue equal to 0.9 and bidder utilities equal to $0.5$ and $0.1$. In this example, (b) has a better performance than (a) because (b) brings higher expected revenue to the exchange.
\end{example}

\begin{figure}[h!] 
\centering 
\includegraphics[width=0.7\textwidth]{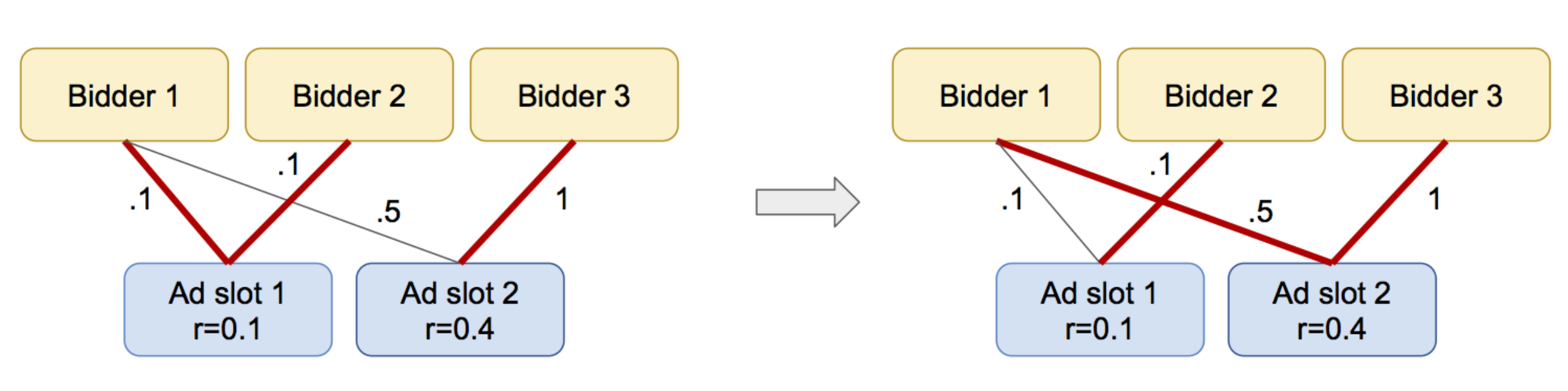}
\caption{Illustration of Example \ref{ex:2}. Callout mechanism (a) on the left and (b) on the right.} 
\label{fig:ex2}
\end{figure}

\begin{example}\label{ex:2}
Consider the following setting (See Figure~\ref{fig:ex2}). Suppose $A =\{a_1, a_2 \}$, $B = \{1,2,3\}$, $q_1= q_2 = q_3 = 1$, and $\delta=1$.
Let's assume $b^1_{1} = 0.1, b^2_{1}=0.5$, $b^1_{2} = 0.1$ and $b^2_{2}=0$, and finally $b^1_{3} = 0$ and $b^2_{3}=1$. Also $r_1= 0.1$ and $r_2= 0.4$.
Both mechanisms (a) and (b) call bidder 2 to item 1 and bidder 3 to item 2. Mechanism (a) calls bidder 1 to item 1, while mechanism (b) calls her to item 2. In this example both callout mechanism (a) and (b) results in a revenue equal to $0.5$ and both clear the market. However, in this example mechanism (2) decreases the utility that bidder 3 earns---the utilities earned by bidders 1 and 2 remain unchanged and equal to 0, while that of bidder 3 decrease from 0.6 to 0.5. 
\end{example}

\subsection{Proof of Proposition~\ref{prop:eq}}
Let us assume that the bidder's initial belief about the exchange's choice of $e$ is such that she chooses to participate in the first stage. 
If from the resulting information set she observes $u^e \geq u$, then the sub-game perfect criterion requires her to participate in the second stage as well. That results in a payoff equal to $c^e(1+\delta)$ for the exchange and $u^e(1+\delta)$ for the bidder.
On the other hand, if the bidder observes a utility of $u^e < u$ after participating in the first stage, then the sub-game perfect criterion requires that she stop participating in the second stage. This choice in turn results in a payoff equal to $c^e$ for the exchange and $(u^e+\delta u)$ for the bidder.

Let $e$ specify the exchange's strategy (i.e. its choice of callout mechanism). Two cases are possible: either $u^e \geq u$ or $u^e < u$. If former is the case, then the exchange's payoff is maximized when $e=e^*$ (by definition), and if latter is the case, then the exchange's payoff is maximized when $e= \arg\max_{e'} c^{e'}$. Therefore when $\max_e c^e \leq c^{e^*} (1+\delta)$ (or equivalently $\frac{\max_e c^e}{c^{e^*}} -1 \leq \delta$) the exchange's best response to the bidder's strategy is to choose $e^*$. \qed

\subsection{The Greedy Algorithm}\label{app:other}
Algorithm~\ref{alg:greedy} describes our baseline, GRA.

\begin{algorithm}
\caption{The Greedy Algorithm (GRA)}
\label{alg:greedy}
\begin{algorithmic}[1]
\State \textbf{Input:}  $K \in \mathbb{N}$, $\epsilon>0$.
\State Start with $t = 0$.
\State Let $\hat{D}_i$ be the estimated bidding distribution for bidder $i$. Start with equal estimations for all bidders.
\While{there exist more ad slots}
\State $t=t+1$.
\State Receive ad slot $a_t$.
\State $B_t = \emptyset$.
\For{$i=1,2,\cdots,K$}
\State Approximate the marginal revenue impact of adding bidder $i$ to $B_t$ using $\frac{1}{\epsilon}$ repetitions.
\State Add that bidder with highest marginal revenue impact to $B_t$.
\EndFor
\State Run the auction among bidders in $B_t$.
\State Update $\hat{D}_i$ for all $i$ using $\vb_{t}$.
\EndWhile
\end{algorithmic}
\end{algorithm}

\subsection{Proof of Proposition~\ref{prop:ShA}}
Let $B_k$ denote the set of bidders with the highest $k$ bids. Consider the case where all bidders submit bids equal to $1$. 
Define $\ve_k = (1,1,..,1,0,..,0)$ to be the vector that places ones in the first $k$ positions, and zeros in the $n-k$ positions thereafter. For $k=2,3,\cdots,n$ properties 1, 2, and 3 can then be translated to:
\begin{enumerate}
\item Symmetry:  there exists $c$ such that $\vR_{B_k} = c \ve_k$. 
\item Linearity: $\vR_{B_k} = \vA \ve_k$ for some fixed matrix $\vA$.
\item Conservation of Revenue:  $\textbf{1}^T \vR_{B_k}= 1$.
\end{enumerate}
From 1 we have that $R_1=\cdots=R_k=c$. Combining this with 3, we have $c = \frac{1}{k}$, or $\vR_{B_k} = \frac{1}{k} \ve_k$. Combining the latter with property 2 we have $\frac{1}{k} \ve_k = \vA \ve_k$ for $k=2,3,\ldots,n$. Similarly we have that $\vA \ve_1 = 0$. This uniquely identifies $\vA$ as follows\footnote{For a second-price auction with reserve price $r$ the derivation is similar. The only difference is that we must treat bids below $r$ as $0$, because such bids cannot have positive revenue impact.}:
$$
\vA = \left( 
\begin{array}{cccccc}
0 & \frac{1}{2} & -{1 \over 6} & \ldots &-{1 \over (n-1)(n-2)} & -{1 \over n(n-1)}\\
0 & \frac{1}{2} & -{1 \over 6} & \ldots & -{1 \over (n-1)(n-2)} & -{1 \over n(n-1)}\\
0 & 0 & +{1 \over 3} & \ldots & -{1 \over (n-1)(n-2)} &-{1 \over n(n-1)}\\
\vdots & \vdots & \vdots & \ldots & \vdots & \vdots\\
0 & 0 & 0 & \ldots & -{1 \over (n-1)(n-2)} & -{1 \over n(n-1)}\\
0 & 0 & 0 & \ldots & +{1 \over (n-1)} & -{1 \over n(n-1)}\\
0 & 0 & 0 & \ldots & 0 & {1 \over n}\\
\end{array} \right)
$$
This finishes the proof.
\qed

\subsection{Running Times}
\begin{table*}[ht!]\label{tab:rt}
\centering
\caption{Running time of each algorithm in terms of the number of bidders $n$, auctions $m$, and accuracy $\epsilon$}
\begin{tabular}{ |l|l|l|l|l|l|l|l|l|l|}
\hline
Algorithm & RQT & GRA & ShA & BAR & WIN & SPD & BID & RVC & RNK\\
\hline
Running Time & $\theta(\frac{nm}{\epsilon})$ & $\theta(\frac{n^2 m}{\epsilon})$ & $\theta(n^2 m)$ & $O(n m)$ & $\theta(m)$ & $\theta(m)$ & $\theta(n m)$ & $\theta(m)$ & $\theta(n \log(n) m)$ \\
\hline
\end{tabular}
\end{table*}